\documentclass[prr,10pt,twocolumn, superscriptaddress,floatfix,showpacs,nofootinbib]{revtex4-2}

\usepackage[sc,osf]{mathpazo}
\usepackage[scaled=0.86]{berasans}  % URL font that go well wtih palatino
\usepackage[colorlinks=true, allcolors=blue, urlcolor=blue]{hyperref}  %Hyperlinks (pink, green, blue)
\usepackage{graphicx} % Package to insert exteral figures
\usepackage{amsmath,mathtools,amssymb,amsthm,bm,amsfonts,mathrsfs,bbm} %Usefull math packages

\usepackage{xspace}  %Useful to add space in macros
\usepackage{pgfplots}
\usepackage{xcolor,colortbl}
\usepackage{array}
\usepackage{bigstrut}
\usepackage{mathrsfs}
\usepackage{dsfont}
\usepackage{quantikz}
\usepackage{multirow}
\usepackage{tikz}
\usepackage{quantikz}
\usepackage{makecell}
\usepackage{caption}
\usepackage{subcaption}
\usepackage{ragged2e}
\usepackage{appendix}

\DeclareCaptionJustification{justified}{\justifying}
\usepackage{tabularx}
\newcolumntype{C}{>{\centering\arraybackslash}X}

\usepackage{lipsum}

\usepackage{verbatim}

\usepackage{comment}
\newcommand{\N}{\mathbb{N}}

\newcommand{\tr}{\text{tr}}

\newcommand{\be}{\begin{equation}}
\newcommand{\ee}{\end{equation}}
\newcommand{\bea}{\begin{eqnarray}}
\newcommand{\eea}{\end{eqnarray}}
\newcommand{\bes}{\begin{equation*}}
\newcommand{\ees}{\end{equation*}}
\newcommand{\beas}{\begin{eqnarray*}}
\newcommand{\eeas}{\end{eqnarray*}}

\newcommand{\I}{\mathds{1}}

\def\M{\widetilde{M}}
\def\N{\widetilde{N}}

\def\tr{\mathrm{tr}}

\begin{document}

%\title{ Proposal for Purifying Noisy Preparation and Measurements of Superconducting Qubits }

\title{ Protocol for Purifying Noisy Preparation and Measurements of Qubits }

%#####################################
\author{Jaemin Kim}
%\email{woals6584@kaist.ac.kr} 
\affiliation{School of Electrical Engineering, Korea Advanced Institute of Science and Technology (KAIST), $291$ Daehak-ro, Yuseong-gu, Daejeon $34141$, Republic of Korea }
\affiliation{Department of Electronic Systems, Aalborg University, 9220 Aalborg, Denmark}

\author{Seungchan Seo}
%\email{ichi9505@kaist.ac.kr}
\affiliation{School of Electrical Engineering, Korea Advanced Institute of Science and Technology (KAIST), $291$ Daehak-ro, Yuseong-gu, Daejeon $34141$, Republic of Korea }

\author{Jiyoung Yun}
%\email{jiyoungyun@kaist.ac.kr}
\affiliation{School of Electrical Engineering, Korea Advanced Institute of Science and Technology (KAIST), $291$ Daehak-ro, Yuseong-gu, Daejeon $34141$, Republic of Korea }

\author{Benjamin Lienhard}
%\email{ }
\affiliation{Technical University of Munich, TUM School of Natural Sciences,
Department of Physics, 85748 Garching, Germany}
\affiliation{Walther-Meißner-Institut, Bayerische Akademie der Wissenschaften, 85748 Garching, Germany}

\author{Joonwoo Bae}
%\email{joonwoo.bae@kaist.ac.kr}
\affiliation{School of Electrical Engineering, Korea Advanced Institute of Science and Technology (KAIST), $291$ Daehak-ro, Yuseong-gu, Daejeon $34141$, Republic of Korea }

%########################################

\begin{abstract}
Noise affecting qubit preparation and measurements accounts for a significant fraction of errors in quantum information processing. This is especially critical in tasks like variational quantum algorithms, quantum error correction, and entanglement distribution through repeaters. In this work, we present a protocol to purify noisy SPAM, effectively suppressing these errors to an arbitrarily low level. For instance, in a realistic scenario where qubits contain error rates around $0.05$ in both preparation and measurement, the protocol can suppress error rates up to $10^{-3}$ with two ancillas and $10^{-6}$ with four ancillas. We show how to distill error-free SPAM by repeating noisy SPAMs. The protocol is also feasible with superconducting qubits. We envisage that our results can be used to realize quantum information tasks in computing and communication with negligible SPAM errors. 
\end{abstract}

\maketitle
%\pacs{03.65.Ud, 02.50.Le, 03.67.Ac}
%########################################

%Realizations of quantum information protocols are currently limited by the inherently noisy components of the processors. 

To address the problem of noise inherent in quantum devices, it is increasingly important to mitigate or suppress errors due to noisy state preparation and measurements (SPAM). For example, hybrid quantum-classical computing approaches---such as variational quantum algorithms, which are promising for near-term quantum technologies---rely on repeated SPAM operations~\cite{RevModPhys.94.015004, TILLY20221}. In doing so, SPAM errors may propagate through hybrid quantum-classical computing. Furthermore, SPAM is an essential element in quantum error-correcting codes~\cite{gottesman1997stabilizercodesquantumerror}. Having sufficiently low SPAM error rates is also significant for fault-tolerant quantum computing. 

{
Significant efforts have been devoted to improving the readout of noisy measurements. The purification of noisy measurements has been proposed by analyzing measurement statistics in an information-theoretic manner, e.g., using maximum-likelihood information principles \cite{DallArno2010purification}. Active readout-error mitigation has been proposed by following advantage distillations or quantum error correction schemes \cite{Hicks2022activereadout, Guenther2022improving, kimyunbae2024}. It is also worth mentioning a recent proposal that attempts to simulate or reproduce other measurements, such as those without noise \cite{LindenSkrzypczyk2025almostperfect}. We remark that all of the aforementioned proposals exploit ancillary qubits prepared in a noiseless manner. It remains unclear whether noisy preparations of ancillary qubits can be used to suppress readout errors. A cost-effective protocol with a minimal number of ancillary qubits is also desired.}

%In fact, little is yet known about how to deal with both preparation and measurement errors on an equal footing.} 

%More recently, Linden and Skrzypczyk~\cite{LindenSkrzypczyk2025almostperfect} studied the general problem of reproducing one quantum measurement using multiple uses of another from a resource-theoretic and measurement-simulation perspective, but their framework is not specialized to SPAM purification and does not provide the concrete resource estimates for near-term implementations.

%proposed active readout-error mitigation by encoding each qubit into a multi-qubit state immediately before measurement, followed by classical decoding such as majority voting or syndrome-based correction. 

%In contrast to these previous works, the protocol presented here treats preparation and measurement noise on equal footing, operates through quantum post-selection rather than decoding alone, and provides a complete analytical framework---including convergence limits under noisy CNOT gates, a closed-form purification threshold, quantified resource requirements, and applications to entanglement distillation and swapping in quantum networks.}

%\cite{DallArno2010purification,Hicks2022activereadout,Guenther2022improving,LindenSkrzypczyk2025almostperfect}

%Among these, }

SPAM errors are due to the inability to realize a fiducial setting. A noiseless scenario with $n$ qubits initialized in a fiducial state is denoted by $|0\rangle\otimes\cdots \otimes |0\rangle=|0\rangle^{\otimes n}$. Similarly, a measurement of each qubit in the computational basis follows as $\{ |0\rangle, |1\rangle \}$~\cite{nielsen_chuang_2010}. Noisy SPAMs lead to a deviation from the noiseless setting. For instance, a noisy initialization of a single qubit can be described by a mixed state $\rho_0=  f |0\rangle \langle 0| + (1-f) |1\rangle \langle 1|$ with $f\in(1/2, 1]$, and suppressing the initialization error can be quantified by the fraction $f$.

In this work, we present a practical and cost-effective protocol for purifying noisy SPAM using noisy SPAM and a set of auxiliary qubits. The protocol is practical in that it works when both preparation and measurements of qubits are noisy, in particular, for error rates of present-day technologies, e.g. \cite{Acharya2025}, and cost-effective since it may efficiently work with even a single ancilla qubit under realistic conditions. We show that, under realistic assumptions, a few qubits are immediately cost-effective to reach an error rate as low as $10^{-3}$. The proposed protocol can suppress SPAM errors to arbitrarily low values. We propose an experimental implementation of the protocol for superconducting qubits, leveraging existing resources already available on state-of-the-art processors, and thus remaining resource-friendly. We then consider the purification of noisy SPAM for key tasks in a quantum network; one is two-way entanglement distillation protocols where local measurements can be noisy, and the other is a measurement for entanglement swapping in a quantum repeater. Thus, our results envisage an effectively noiseless quantum network with noisy SPAM.

{\it SPAM errors}. Throughout this work, we define a noiseless scenario as one in which a quantum system is initialized in state $|0\rangle$, and measurements are performed in the computational basis $\{M_0,M_1\}$, where $M_k = |k \rangle\langle k |$ for $k =0,1$. We denote a noisy state preparation by the density matrix $\rho$, given by: 
\bea
\rho =
\begin{bmatrix}
\rho_{00} & \rho_{01} \\
\rho_{10} & \rho_{11}.
\end{bmatrix} \label{eq:f}
\eea
The fidelity of the preparation is quantified by $f = \langle 0| \rho | 0\rangle$. Similarly, we describe noisy measurements using positive operator-valued measure (POVM) elements $\widetilde{M}_{k}$ corresponding to outcomes $k\in \{ 0,1\}$. The measurement noise is characterized by a noise fraction $q =  \langle \bar{k}| \widetilde{M}_k | \bar{k}\rangle$, where $\bar{k}=k \oplus1$ denotes bitwise addition. 
Throughout, we assume $q \in [0, 1/2)$, corresponding to a measurement that performs better than random guessing. A measurement with $q > 1/2$ can always be converted to one with $q < 1/2$ by relabeling outcomes.

Without loss of generality, we assume that noisy POVM elements are diagonal as follows, 
\bea
\widetilde{M}_k   = (1-q) M_k + q M_{k\oplus 1}~~\mathrm{for}~~ k=0,1. \label{eq:nm}
\eea
This is because one can diagonalize a POVM element by incorporating random Pauli-$Z$ gates as follows. For a qubit POVM element $\hat{M}_{k}$ having a noise fraction $q$, it holds that, 
\bea
\frac{1}{2} \left( \hat{M}_{k} + Z \hat{M}_{k} Z \right) = (1-q) M_k + q M_{ k\oplus 1}~~\mathrm{for}~~ k=0,1. \nonumber
\eea 
The left-hand side above is equivalent to cases where one randomly applies a Pauli-$Z$ gate to a state before a measurement is performed. A POVM element resulting from random applications of a Pauli-$Z$ gate is diagonalized while the noise fraction $q$ remains unaltered. That is, a POVM element can be generally diagonalized, and it suffices to consider noisy POVMs in Eq. (\ref{eq:nm}). 

\begin{figure}[t]
    \centering
    \includegraphics[width=.49 \textwidth]{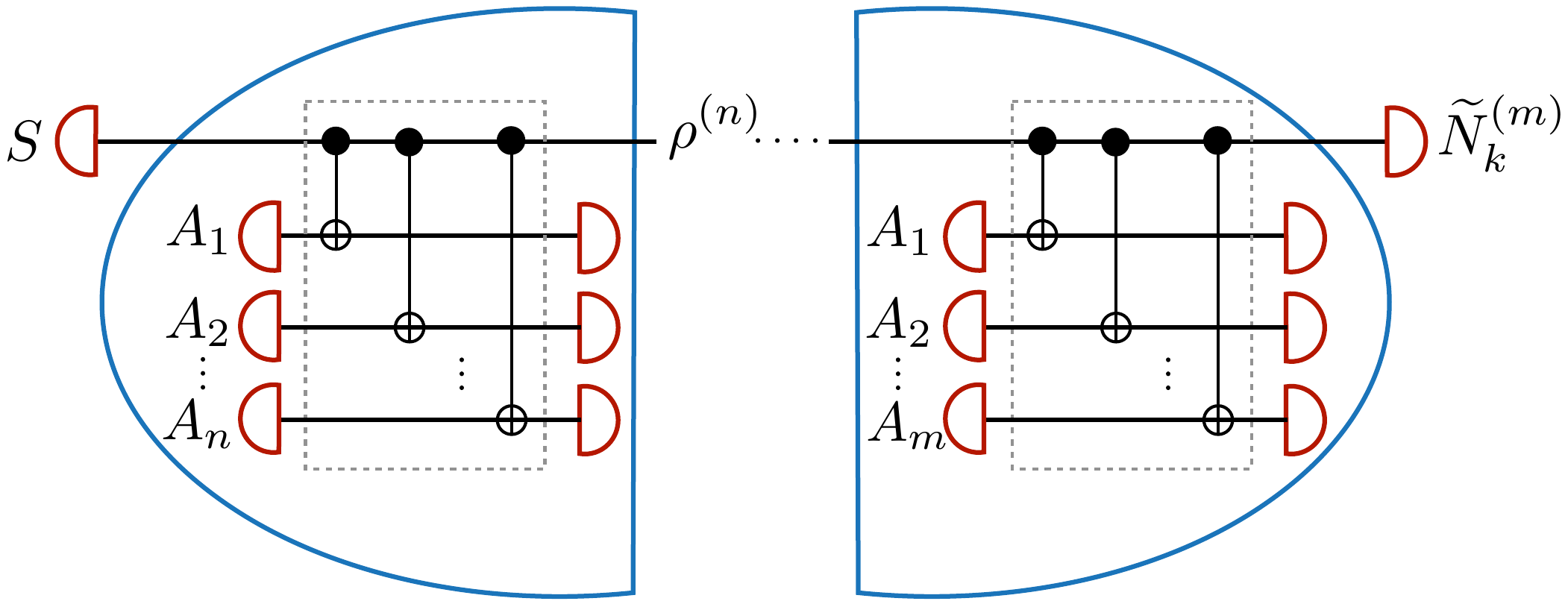}
    \caption{Repeated noisy SPAM can purify noisy SPAM. The protocol starts with a collective CNOT applied to noisy qubits $SA_{1}\cdots A_n$. The resulting state $\rho^{(n)}$ is accepted once the outcomes of the noisy qubit measurements $A_1\cdots A_n$ are all zeros $0^n$. The fidelity converges to $1$ as $n$ increases. A noisy measurement on a system $S$ is purified with noisy SPAM with $m$ target qubits; a purified POVM element giving an outcome $k$ is denoted by $\N_{k}^{(m)}$, having a noise fraction converging to $0$. }
  \label{fig:scenario}
\end{figure}

{\it Protocol for purifying qubit-state preparation.} We propose a protocol for purifying noisy qubit-state preparations, with the goal of producing an initial state whose fidelity $f$ approaches unity arbitrarily closely. The protocol requires $n$ ancillary qubits and a collective CNOT-gate on $(n+1)$ qubits, defined as: 
\bea
V_{n+1}^{SA_1\cdots A_{n }} = |0\rangle \langle 0|^S \otimes \I^{\otimes  n  } + |1 \rangle \langle 1|^S \otimes X^{\otimes  n  } \label{eq:cv}
\eea
where $X$ denotes a Pauli-$X$ gate. This collective operation $V_n$ can be implemented as a sequence of standard two-qubit CNOT gates, $V_{n+1}^{SA_1\cdots A_{n }} =  \Pi_{i=1}^n V_{2}^{SA_i}$, see Fig.~\ref{fig:scenario}, where $V_{2}^{SA_i}$ is a CNOT gate with the system qubit $S$ as control and the auxiliary qubit $A_i$ as target. Following the application of the collective CNOT gates, noisy measurements are performed on $n$ target qubits, Eq.~(\ref{eq:nm}). The resulting state in the control register is accepted, denoted by $\rho^{(n)}$, only if all $n$ measurement outcomes of the target qubits are zeros, meaning the output string is $0^{n}$. It can be shown that the fidelity of the post-selected state satisfies $ f^{(n)} = \langle 0| \rho^{(n)} |0\rangle \longrightarrow 1$, demonstrating that the protocol enables arbitrarily accurate state preparation. A detailed derivation is provided in Appendix \ref{app:a}. \\

{\it {\bf Proposition 1.} For all $\epsilon>0$, there exists $N\ge 1$ such that $|f^{(n)} -1|< \epsilon $ whenever $n>N$.}\\

{\it Protocol for purifying noisy measurements.} The purification protocol of noisy qubit measurements aims to approximate the results of noiseless measurements in a computational basis. We show that the protocol can purify a noisy measurement with POVM elements in Eq.~(\ref{eq:nm}) by suppressing an error rate $q$. 

The purification protocol, depicted in Fig.~\ref{fig:scenario}, comprises noisy qubit-state preparations $\rho$ and noisy qubit-state measurements in Eq.~(\ref{eq:nm}). To purify a noisy measurement of state $\sigma$, $m$ ancillary qubits are initialized in $\rho^{\otimes m}$. A collective CNOT $V_m$ is applied to $\sigma\otimes \rho^{\otimes m}$. Subsequently, measurements are performed on each of the $(m+1)$ qubits individually, as described in Eq.~(\ref{eq:nm}). The results are accepted only if the measurement results are identical, either $0^{m+1}$ or $1^{m+1}$.

Suppose that the measurements are identical and accepted. We write that $\widetilde{N}_{k}^{(m)}$ is a POVM element that yields $k$ in the first register, since the results $m+1$ are identical. The noisy fraction for the POVM element is given by $q^{(m)} = \langle \bar{k}| \widetilde{N}_{k}^{(m)} | \bar{k}\rangle$ and it turns out that $q^{(m)}\rightarrow 0$ as $m$ increases. The derivation is shown in Appendix \ref{app:b}.  
\\

{\it {\bf Proposition 2.} For all $\epsilon>0$, there exists $N\geq 1$ such that $q^{(m)}<\epsilon$ whenever $m>N$.}\\

The results in Propositions~1 and 2 extend to cases where ancilla qubits are prepared with different values of preparation fidelity $f_i$ and measurement noise $q_i$. The convergence for the purification, $f^{(n)} \to  1$ and $q^{(m)} \to 0$, holds whenever the noise parameters satisfy $f_i > 1/2$ and $q_i < 1/2$ for each $i$. The proof is detailed in Appendix \ref{app:b}.

\begin{table}[t]
\label{tab:1}
\caption{  Purification of noisy measurements with noisy SPAM is demonstrated. Note that $T^{(n)}$, see Eq. (\ref{eq:pud}), is a trace distance between a noiseless POVM element and a purified one with $n$ ancilla qubits, and $p_{s}^{(n)}$ denotes the success probability.}
\label{tab:1}
\centering
\begin{tabular}{|c|c||c|c||c|c|c|}
\hline
$\mathrm{error ~rate}$ & $~~T^{(0)}~~$ &  $~~T^{(1)}~~$ & $~~p_{s}^{(1)}~~$ & $~~ T^{(2)}~~$ & $~~p_{s}^{(2)}~~$ \\
\hline \hline
$1-f=q=0.1 $ & \makecell{$0.1$} &  \makecell{$0.024$ } & \makecell{$0.756$} 
& \makecell{$0.005$}  & \makecell{$0.608$} \\
\hline
$1-f=q=0.05 $ & \makecell{$0.05$ } 
& \makecell{$0.005$} & \makecell{$0.864$} 
& \makecell{$0.001$}  & \makecell{$0.779$} \\
\hline
\end{tabular}
\end{table}

We note that the purification protocol is distinct from {\it quantum error mitigation}~\cite{PhysRevX.8.031027, RevModPhys.95.045005} in the following sense. The error-mitigation protocols aim to estimate noiseless expectation values of observables by repeating experiments. The proposed protocol attempts to purify noisy SPAM and aims to achieve noiseless statistics. As for shortcomings, the mitigation technique mentioned above may suffer from the costs of sampling~\cite{PhysRevLett.131.210602}, whereas the proposed purification protocol may have low success probabilities since it relies on post-selection. In Table \ref{tab:1}, we consider a realistic scenario where SPAM errors are about $5$-$10$ \%, and show that one or two ancillary qubits are sufficient to reach an error rate up to $10^{-3}$. Note also that success probabilities are higher than $0.5$. Further analysis with realistic parameters is provided in Appendix \ref{app:c}. In fact, in what follows, we show that the resource requirement to suppress SPAM errors up to a desired level, for example, $10^{-3}$ or even less, can be achieved with a few ancilla qubits.

{\it Realistic scenarios.} To quantify a purification of noisy measurements, we use the trace distance, $T(A,B) = \frac{1}{2}\| A-B \|_1$ for Hermitian operators $A$ and $B$ where $\| X \|_1 = \tr\sqrt{X^{\dagger} X}$. 

For a purified POVM element $\widetilde{N}_{k}^{(m)}$ with $m$ additional qubits, the quantification may be addressed for $\delta >0$ such that 
\bea
T^{(m)}   : = T ( \N_{i}^{(m)} , M_i ) \leq  \delta~~\mathrm{for}~~i=0,1.  \label{eq:pud}
\eea
Note also that the distance above corresponds to the noise fraction, that is, $q^{(m)} = T^{(m)}$. Equivalently, one can find the fidelity $ F(\N_{i}^{(m)} , M_i ) =1-  T^{(m)}  \geq 1-\delta$ since $M_i$ to be rank $1$. The other parameter is the probability of success that the measurement outcomes are identical, which can be computed as $ p_{succ}^{(m)}= \alpha^m (1-q) + (1-\alpha)^m q$, where $\alpha = f(1-q) + (1-f)q$.

State-of-the-art quantum processors suffer from SPAM errors on the order of $1$-$10\%$~\cite{Acharya2025, Bravyi2024}. Assuming that error rates are balanced, that is, $1$-$f = q \approx 5\%$, we find that an error rate is suppressed up to $10^{-3}$, i.e., $1$- $f^{(n)} = q^{(n)} < 10^{-3}$ for $n \geq 2$. The probability that three measurements are identical, that is, $n=2$, is $0.7785$, see Table \ref{tab:1}. Therefore, the protocol with a few ancillary qubits is immediately efficient and cost-effective. More instances of the resource requirements of the purification protocol are presented in Appendix \ref{app:c}.

We note that the balanced error rates are assumed for the purpose of demonstrating the purification of noisy resources. We recall that the convergence in the purification is shown for general values of $f$ and $q$ in the propositions above. More instances when the parameters are not balanced, i.e., $1-f\neq q$, are demonstrated in Appendix \ref{app:d}.

{\it Experimental feasibility.} To demonstrate the feasibility of the proposed protocol, we present a blueprint tailored to state-of-the-art superconducting quantum processors that is resource-efficient and readily implementable. Superconducting qubits are among the leading platforms for the realization of scalable quantum information processors~\cite{annurev:/content/journals/10.1146/annurev-conmatphys-031119-050605, Acharya2025, Bravyi2024}. Despite advances in superconducting qubit architectures, the fidelity of qubit-state preparation is often constrained by residual excitations, ground-state heating, and imperfect gate operations. More importantly, readout errors remain a significant challenge. These errors---caused by residual coupling, crosstalk, or limited signal-to-noise ratios---can misidentify the state of a qubit during measurement~\cite{PhysRevApplied.17.014024}.

\begin{figure}[t]
    \centering
    \includegraphics[width=.49 \textwidth]{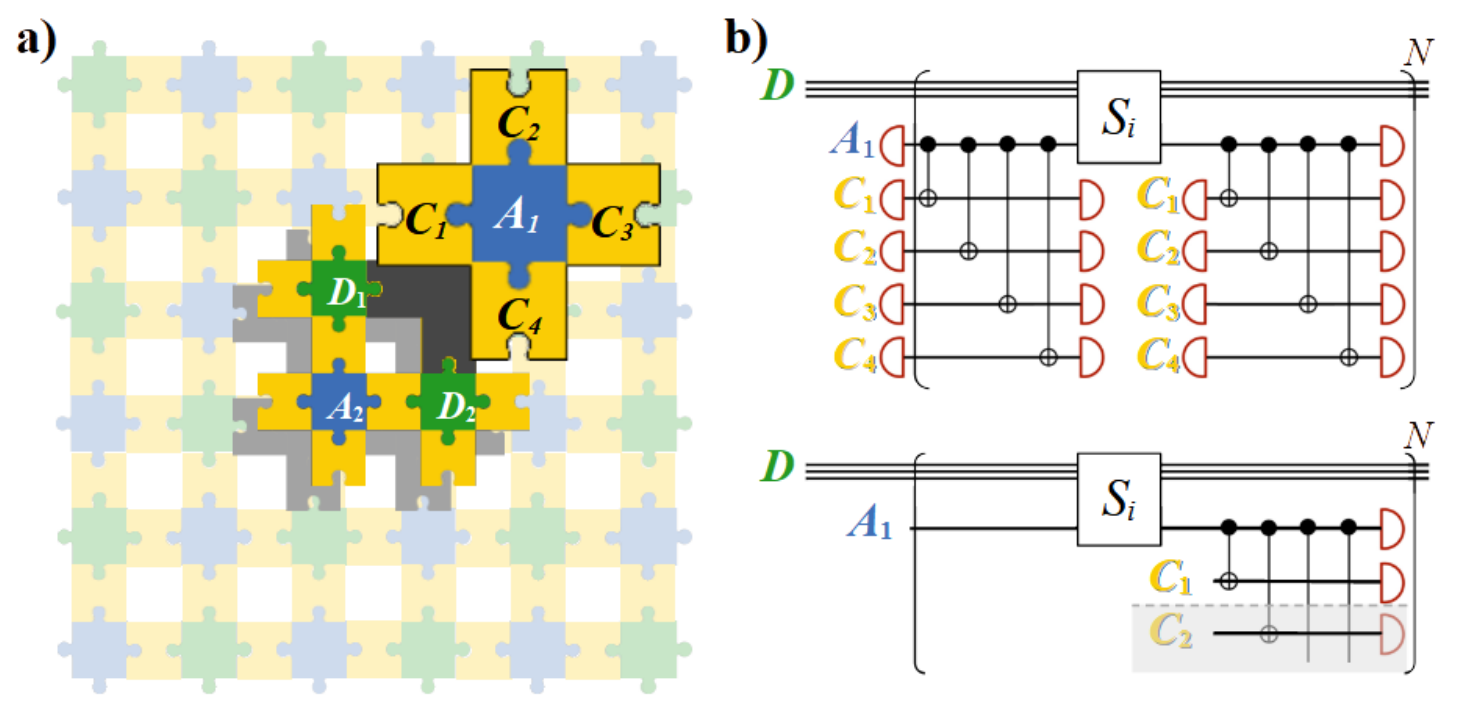}\vspace{-2mm}
    \caption{SPAM purification protocol on a superconducting quantum processor encoded with surface code. a) Data qubits (D, green), ancilla qubits (A, blue), and tunable couplers (C, yellow) are represented as puzzle pieces. b) Top: Full protocol implementation using all four tunable couplers as ancilla qubits for each QEC cycle $i$ (up to $N$ cycles) and corresponding syndrome measurement $S$. Bottom: Resource-efficient implementation using one (or two) couplers, reducing SPAM error from 1.3\% to 0.05\% (0.002\%).}
  \label{fig:superconducting}
\end{figure}

Modern superconducting architectures frequently incorporate tunable couplers between qubits, which serve to minimize crosstalk and facilitate interactions between two qubits on demand~\cite{PhysRevApplied.10.054062, Acharya2025, PhysRevLett.127.080505}. These features make superconducting qubits particularly well-suited for implementing the proposed SPAM purification protocol without requiring additional hardware resources. Tunable couplers are qubits themselves and can thus act as auxiliary qubits needed to implement the proposed SPAM purification protocol, as illustrated in Fig.~\ref{fig:superconducting}.

In superconducting qubit systems, the surface code---currently the most widely implemented quantum error correction scheme---relies critically on minimizing SPAM errors to preserve the integrity of logical qubits~\cite{PhysRevA.86.032324,Acharya2025}. Each plaquette typically uses a dedicated ancilla qubit. Each ancilla qubit is surrounded by four tunable couplers enabling on-demand connection to respective data qubits. 

We propose repurposing tunable couplers as ancillary qubits for the presented SPAM purification protocol. Each data or ancilla qubit can access up to four couplers. Utilizing couplers as qubits requires calibrating a two-qubit gate that enables conditional phase accumulation specific to the target qubit, while the phase on the untargeted qubit cancels out. This calibration involves optimizing the coupler’s operating frequency, pulse duration, and shape. Control of the tunable coupler can be achieved either via an additional dedicated control line or through residual coupling to nearby control lines.

Readout can be performed via an additional resonator or through parasitic coupling to the resonator of the associated data or ancilla qubit. The coupler’s tunability allows reducing the frequency gap between its operating point and the resonator, compensating for weaker coupling.

This protocol temporarily reassigns couplers as qubits to support SPAM purification. While it requires extra calibration for gate tuning and parallel readout, it avoids additional hardware and is compatible with current superconducting quantum processors.

To estimate the utility of the proposed protocol, we adopt performance metrics comparable to state-of-the-art quantum processors~\cite{Acharya2025}, as summarized in Tab.~\ref{tab:chip}. 
%Based on these metrics, our protocol can be implemented with the capabilities of current quantum hardware.

Ancilla-qubit initialization achieves significantly higher fidelity than readout, making SPAM errors predominantly due to readout. Given the relatively long readout and signal processing times, the proposed implementation focuses on improving readout fidelity, with overhead limited to the number of CNOT gates between the tunable coupler and the ancilla. 

The ancilla-SPAM error is approximately 1.3\%. Using a tunable coupler during measurement---requiring one CNOT gate and incurring $\sim$0.1\% decoherence---this can be reduced to 0.05\% with over 95\% acceptance. With two couplers, the SPAM error drops to 0.002\%, with $\sim$0.2\% decoherence and 92\% acceptance. Given access to four couplers, we can select the most suitable one or two, achieving substantial SPAM suppression while maintaining high acceptance.

% \begin{table}[]
%     \centering
%     \caption{Superconducting quantum processor performance metrics.}
%     \begin{tabular}{l p{1cm} r}
%          Readout error		    && 0.01	\\
%          Readout duration	    && 400~ns\\
%          Coupler readout error   && 0.02 \\
%          CNOT-gate error	        && 0.01	\\
%          CNOT-gate duration	    && 100~ns\\
%          $T_1$			        && 70~$\mu$s\\
%          $T_{2, \rm CPMG}$		&& 85~$\mu$s\\
%          FPGA signal processing	&& 400~ns\\
%          QNDness			        && 0.001\\	
%          Reset error			    && 0.002\\	
%     \end{tabular}
%     \label{tab:chip}
% \end{table}

\begin{table}[t]
    \centering
    \caption{Superconducting quantum processor performance metrics.}
    \begin{tabular}{l@{\hspace{1cm}}r}
         Readout error            & 0.01 \\
         Readout duration         & 400~ns\\
         Coupler readout error    & 0.02 \\
         CNOT-gate error          & 0.01 \\
         CNOT-gate duration       & 100~ns\\
         $T_1$                    & 70~$\mu$s\\
         $T_{2, \rm CPMG}$        & 85~$\mu$s\\
         FPGA signal processing   & 400~ns\\
         QNDness                  & 0.001\\
         Reset error              & 0.002\\
    \end{tabular}
    \label{tab:chip}
\end{table}

{\it Purifying SPAM errors with noisy gates.} We reiterate that the purification protocol needs noisy SPAM and noiseless CNOT gates. Here, we relax these conditions and consider the purification protocol with imperfect CNOT gates~\cite{PhysRevLett.83.4200}, 
\bea
V_2 (~\cdot~) V_{2}^{\dagger}\mapsto (1- \epsilon) V_2 (~\cdot~) V_{2}^{\dagger} +   \epsilon~\tr[~\cdot~] \frac{\I}{2} \otimes \frac{\I}{2},~ \label{eq:noisecnot}
\eea
where $\epsilon$ represents a noise fraction. It turns out that noisy SPAM cannot be completely purified with noisy CNOT gates. With three noise parameters, $1-f$, $\epsilon$, and $q$, whthe protocol can reach the fidelity in state preparation and a noise fraction in measurements as follows,
\bea
f_{\epsilon}^{(n)}&&   ~~  \xrightarrow{ n \rightarrow \infty }~~ \frac{1}{1-D+\sqrt{D^2 +1}}< 1 ~~\mathrm{and}~~\label{eq:fep} \\
q_{\epsilon}^{(m)}&& ~~  \xrightarrow{ m \rightarrow \infty }~~ \frac{1}{1+D+\sqrt{D^2 +1}} > 0   ~ \label{eq:mep} 
\eea
where $D= 2(2f -1)(1-2q) (1-\epsilon) /\epsilon$. The derivations are shown in Appendix \ref{app:e}. The estimation of the parameters is also shown in Appendix \ref{app:g}. It can be shown that the qubit-state-preparation fidelity is strictly less than $1$. Furthermore, noisy measurements can be purified up to an error $q_{\epsilon}^{(m)}$, which is strictly positive.

\begin{figure}[t]
    \centering
    \includegraphics[width=.5 \textwidth]{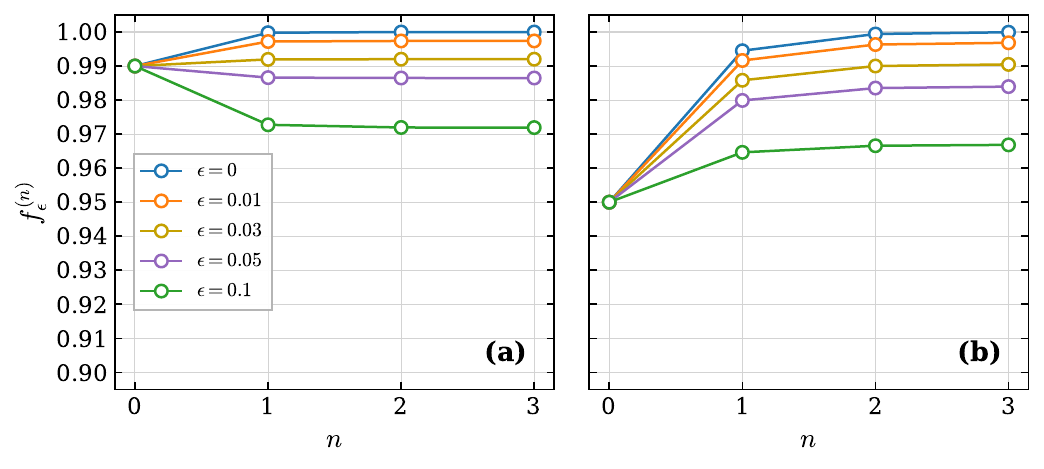}
    \caption{ The protocol purifies noisy qubit-state preparation with resources of noisy SPAM and imperfect CNOT gates on $n$ additional qubits. For noiseless CNOT gates, one can purify noisy preparation up to a fidelity of $1$. For an initial fidelity $f=0.99$, one additional qubit suffices to achieve a noiseless preparation. If a CNOT gate is noisy with an error rate $\epsilon > 0.0374$, the purification is not fulfilled, and, thus, the fidelity decreases (purple and green plots). For $f=0.95$, two or three additional noisy qubits can efficiently increase the fidelity.  }
  \label{fig:graph}
\end{figure}

The purification protocol achieves a fidelity up to $0.984$ with two target qubits for a realistic scenario with balanced error rates $\epsilon=1-f=q=0.05$. The protocol purifies state preparation from $0.95$ up to $0.984$. The success probability with two target qubits is given by {$p_{succ}^{(2)}=0.7357$}. For a noisy measurement, the protocol suppresses an error rate of $0.05$ up to $0.016$, as illustrated in Fig.~\ref{fig:graph}. 

{\it The purification condition.} Eqs.~(\ref{eq:fep}) and (\ref{eq:mep}) show that the convergence depends on the parameters $f$, $q$, and $\epsilon$. This means that the purification protocol does not work if the first iteration does not improve fidelity, defining the condition for purifying noisy SPAM, 
\bea
\mathrm{the ~ purification~condition~}: f < f_{\epsilon}^{(1)}. \label{aeq:pc}
\eea
As an instance, assuming that SPAM errors are balanced, that is, $1-f=q$, an initial fidelity increases by the purification protocol whenever $\epsilon <  \epsilon_c$ 
where the threshold $\epsilon_c$ is given by,
\bea
 \epsilon_c & = &  \frac{  8 f^3 -12f^2 +4f }{ 8 f^3 -12 f^2 +4f -1} ~~~~~\label{eq:ec} \\
 & \approx & 4(1-f) - 28(1-f)^2 + O((1-f)^3). \nonumber
 \eea
The expansion above shows that for error rates $1-f=q$ around $10^{-2}$, a threshold of an error rate $\epsilon_c$ is about four times more relaxed.
In Tab.~\ref{tableIII}, thresholds $\epsilon_c$ are computed for error rates relevant to current state-of-the-art architectures and applications. For example, for an error rate $1-f=q=0.01$, the critical error rate $\epsilon_c$ results as $0.0374$, see Fig. \ref{fig:graph}.

\begin{table}[h]
\caption{ Critical error rates $\epsilon_c$ of a CNOT gate for purifying noisy SPAM are shown, see Eq.~(\ref{eq:ec}).}
\label{tableIII}
\begin{tabular}{c||cccccccc}
\hline
\hline
~$1-f$ ~&& 0~& 0.01 ~&0.03~ & 0.05~ & 0.07~ & 0.1  \\  \hline
~ $\epsilon_c$ ~&& 0~& 0.0374 ~ & 0.0986~ & 0.1460~ & 0.1830~ & 0.2236 \\\hline
 \hline
\end{tabular}
\end{table}

%%%%%%%%%%%%%%%%

{\it Application 1: Distilling entanglement.} Entanglement is a general resource for quantum information applications~\cite{RevModPhys.81.865}. Distilling an entangled bit (ebit),
\bea
|\phi^+\rangle =\frac{|00\rangle +|11\rangle}{\sqrt{2}} \label{eq:ebit}
\eea
is a key task that enables various quantum information tasks, such as entanglement distillation~\cite{kimyunbae2024}. Entanglement distillation begins when two parties share copies of an isotropic state, 
\bea
\rho_F = F|\phi^+\rangle \langle \phi^+| + \frac{1-F}{3}\left( \I \otimes \I- |\phi^+\rangle \langle \phi^+|\right), \nonumber
\eea
which is entangled if and only if the fidelity satisfies $F>1/2$. Two parties consider two copies and apply CNOT gates locally. Performing measurements in the second register, they accept a remaining state in the first register if the measurement outcomes are identical. Then, two parties may share a state with fidelity $F^{'}$, for which we have $F^{'}>F$ whenever $F>1/2$. Hence, all entangled two-qubit states are distillable. 

In Ref.~\cite{PhysRevA.59.169}, it is shown that if the measurements in the second register are noisy, the distillation protocol cannot distill ebits from weakly entangled states. Here, we apply the purification protocol for distilling entanglement with noisy measurements and investigate the resources needed. We write by $\widetilde{M}_{k}^{(n)}$ a POVM element purified with $n$ ancillary qubits; it is obtained by noisy initialization $\rho$ and noisy measurements $\widetilde{M}_k$. We find a recurrence relation in POVM elements in the purification protocol, 
\bea
\M_{k}^{(n )} & = & \tr_{ A_{n} } [ \rho^{A_{n}}    V_{2}^{  \dagger}   ~    \M_{k}^{(n-1) } \otimes  \M_{k}^{ A_{n} } ~ V_{2}^{ } ] ~~\label{eq:arem}
\eea
where noisy SPAMs are only exploited. We also write 
\bea
r_0^{(n)} = \langle k| {\M_k^{(n)}} |k\rangle~~\mathrm{and}~~ r_1^{(n)} = \langle {\bar k}| {\M_k^{(n)}} | {\bar k}\rangle.\label{eq:rrr}
\eea
Alice and Bob apply purified measurements to the distillation of entanglement and may find that 
\bea
F^{'} = \frac{F^2 +  \left( \frac{1-F}{3} \right)^2  + g^{{(n)}}(F) }{ F^2 + 2F \left( \frac{1-F}{3} \right) +  5 \left( \frac{1-F}{3} \right)^2+  4g^{{(n)}}(F)  } \label{eq:FF}
\eea
where
\bea
g^{{(n)}}(F)  =  \left(   \frac{F(1-F)}{3} + \left( \frac{1-F}{3} \right)^2  \right) \frac{r_{odd}^{(n)}}{r_{even}^{(n)}} \nonumber
\eea
with $r_{odd}^{(n)} = 2 r_{0}^{(n)} r_{1}^{(n)}$ and $r_{even}^{(n)} = r_{0}^{(n)}r_{0}^{(n)} + r_{1}^{(n)}r_{1}^{(n)}$. 

Then, distilling entanglement works, i.e., $ F^{'}>F$, for $F> L^{(n)}$, where
\bea
L^{(n)}  = \frac{1}{2} \left(\frac{r_{0}^{(n)} + r_{1}^{(n)}  }{r_{0}^{(n)} - r_{1}^{(n)} } \right)^2. \nonumber%~~ \xrightarrow{n \rightarrow \infty} ~~ \frac{1}{2} \nonumber
\eea
In what follows, we investigate the lower bound $L^{(n)}$ for distillation. Firstly, if the CNOT gates in the purification are noiseless, $\epsilon=0$, we solve the recurrence relations in Eq.~(\ref{eq:rrr}): $r_0^{(n)} = (1-q) \alpha^n$ and $ r_1^{(n)} = q (1-\alpha)^n$ where $\alpha = f(1-q) + (1-f)q$. It follows that $L^{(n)} \xrightarrow{n \rightarrow \infty} \frac{1}{2}$, meaning that an ebit can be distilled from all entangled two-qubit states. Secondly, if a CNOT gate is noisy, $\epsilon>0$, we end up with $ {r_{1}^{(n)}} / {r_{0}^{(n)}} \rightarrow d= \sqrt{D^2 +1} -D$, where $D= 2(2f -1)(1-2q) (1-\epsilon) /\epsilon$. It turns out that 
\bea
L^{(n)} ~~ \xrightarrow{n \rightarrow \infty} ~~   \frac{1}{2} \left( \frac{1 + d }{1 -d} \right)^2 \label{eq:convv}
\eea
which is strictly greater than $1/2$. Hence, weakly entangled states cannot be distilled. Detailed derivations are shown in Appendix \ref{app:h}.

{\it Application 2: Entanglement swapping}. We consider a repeater $R=R_1R_2$ sharing ebits with Alice and Bob, $|\phi^+\rangle^{AR_1}|\phi^+\rangle^{ R_2B}$. A noiseless Bell measurement is desired, where a projection onto an ebit $|\phi^+\rangle^{R_1R_2}$ enables two parties, Alice and Bob, to share an ebit. 
A successful projection for a state $\rho^{AR_1R_2B}$, denoted by a success probability $p_R$, can be realized as follows,
\bea
p_{R}\rho^{AB} = \tr_{R_1R_2}[ \rho^{AR_1R_2B}  V_2 H^{R_1} |00\rangle \langle 00|^{R_1R_2 }   H^{R_1} V_{2}^{\dagger}] \label{eq:sb}
\eea
since $|\phi^+\rangle = V_2 H^{R_1} |00\rangle$ where $H^{R_1}$ denotes a Hadamard transform on $R_1$ and $V_2$ acts on $R_1R_2$. The result $00$ corresponds to a successful projection. Noisy measurements may fail in creating an ebit between two parties,
\bea
\widetilde{p}_{R} \widetilde{\rho}^{AB}  =\tr_{R_1R_2}[ \rho^{AR_1R_2B}  V_2 H^{ R_1 }  \M_{0}\otimes \M_0 H^{ R_1 } V_{2}^{\dagger}],  \label{eq:sig}
\eea
the result of this leads to a mixture of four Bell states between Alice and Bob.

We apply a purified measurement with $m$ noisy additional qubits on $R_1$ and $R_2$, respectively, constructs POVM elements $\M_{0}^{(m)} \otimes \M_{0}^{(m)}$. The resulting state of Alice and Bob after the projection is denoted by $\sigma^{(m)}$ and can be quantified by the fidelity, 
\bea
\widetilde{F}^{(m)} = \langle \phi^+| \sigma^{(m)} |\phi^+\rangle   = \left[ 1+ \left( \frac{1-\alpha}{\alpha} \right)^m \left(\frac{q}{1-q}\right)   \right]^{-2}\nonumber
%=  \left( \frac{\alpha^m (1-q) }{\alpha^m (1-q) + (1-\alpha)^m q} \right)^2. \nonumber
\eea
Since $\alpha>1/2$, one can find $m\geq 1$ such that $\widetilde{F}^{(m)}$ is sufficiently close to $1$.

{\it Conclusion}. In conclusion, we have presented a protocol that purifies noisy SPAM. The protocol exploits noisy SPAM of auxiliary qubits as a resource. It efficiently supports quantum information tasks with noisy SPAM in general and is feasible with currently available quantum technologies, such as superconducting quantum processors. In particular, the protocol enables entanglement distillation with noisy SPAM~\cite{PhysRevA.59.169} as well as entanglement swapping: both are essential to build a quantum network. We have shown that under realistic assumptions, a few qubits are cost-effective in reaching a sufficiently low error rate in SPAM. The proposed protocol offers a general approach to mitigating SPAM errors, which have posed a significant obstacle to the reliable implementation of noisy-intermediate-scale-quantum algorithms, quantum communication protocols, quantum metrology, e.g., \cite{long2022entanglement}, and the performance of quantum error correction codes.

\section*{Acknowledgement}

J.K., S.S., J.Y., and J.B. were supported by the NRF Korea (RS-2024-00408613) and the IITP Korea (RS-2023-00229524, RS-2025-02304540, RS-2025-25464876, RS-2025-25464616). B.L. acknowledges support from the German Federal Ministry of Research, Technology and Space (BMFTR) through the program EQuIPS (Grant No. 13N17232). J.K. was supported, in part, by the Danish National Research Foundation(DNRF), through the Center of Excellence CLASSIQUE, grant nr. DNRF187.

\bibliography{reference}

\newpage
\newpage

\appendix

\section{Purification of Noisy State Preparation}
\label{app:a}
 
\begin{figure}[t]
    \centering
    \includegraphics[width=.49 \textwidth]{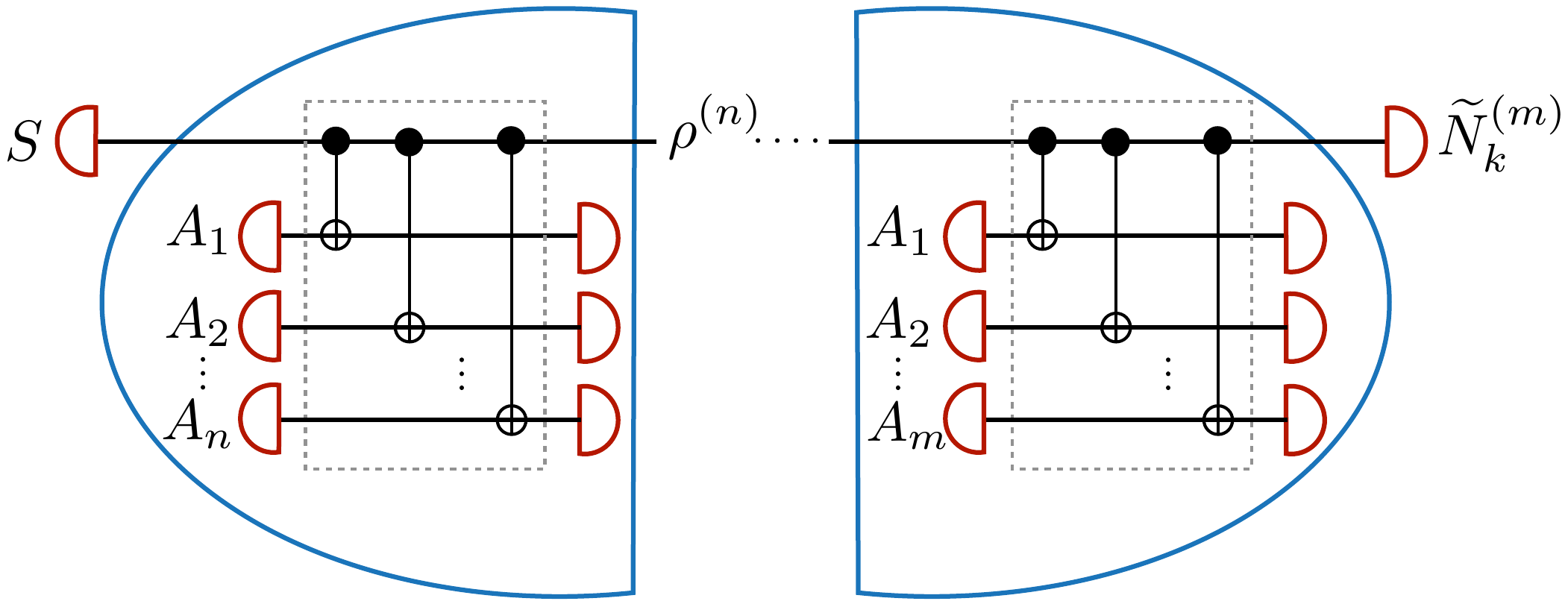}
    \caption{   Noisy SPAM can purify noisy SPAM. A collective CNOT in Eq. (\ref{aeq:cv}) applies to noisy qubits $SA_{1}\cdots A_n$. A resulting state $\rho^{(n)}$ accepted once the outcomes of noisy measurements on noisy qubits $A_1\cdots A_n$ are all zeros $0^n$ has a fidelity converging to $1$ as $n$ increases. A noisy measurement on a system $S$ is purified with noisy SPAM with $m$ target qubits; a purified POVM element giving an outcome $k$ is denoted by $\N_{k}^{(m)}$ having a noise fraction converging to $0$. }
  \label{fig:ascenario}
\end{figure}

{\it Summary of the protocol.} The protocol for purifying noisy state preparation works with $n$ ancillary qubits and a collective CNOT gate on $(n+1)$ qubits,
\bea
V_{n+1}^{SA_1\cdots A_{n }} = |0\rangle \langle 0|^S \otimes \I^{\otimes  n  } + |1 \rangle \langle 1|^S \otimes X^{\otimes  n  } \label{aeq:cv}
\eea
where $X$ denotes a Pauli-$X$ gate. Note also that $V_{n+1}$ can be implemented as a sequence of CNOT gates, $V_{n+1}^{SA_1\cdots A_{n }} =  \Pi_{i=1}^n V_{2}^{SA_i}$, see also Fig. \ref{fig:ascenario}, where $V_{2}^{SA_i}$ is a CNOT gate. After the collective CNOT gates, noisy measurements are performed on $n$ target qubits. A resulting state in the control register is accepted, denoted by $\rho^{(n)}$, only when all $n$ outcomes on target qubits are zeros, i.e., $0^{n}$.

 \bigskip
{\it {\bf Proposition 1.} For all $\epsilon>0$, there exists $N\ge 1$ such that $|f^{(n)} -1|< \epsilon $ whenever $n>N$.}\\
\bigskip

{\it Proof.} We write by $R^{(n)}$ an unnormalized state accepted after a collective CNOT $V_{n+1}$ over a control and $n$ target qubits,  
\bea
R^{(n)} =  \begin{bmatrix}
R^{(n)}_{00} & R^{(n)}_{01} \\
R^{(n)}_{10} & R^{(n)}_{11} 
\end{bmatrix}. \label{aeq:auns}
\eea
Note that the probability that a control qubit is accepted is given by $p_{\rm succ}^{(n)} = \tr R^{(n)}$, and the fidelity is thus as follows, 
\bea
f^{(n)} =    \frac{  R_{00}^{(n) }}{ R_{00}^{(n) }+ R_{11}^{(n) }  }=    \left( 1 +  \frac{ R_{11}^{(n) }}{R_{00}^{(n) }}  \right)^{-1}.  \label{aeq:afn}
\eea
A collective CNOT $V_{n+1}$ can be decomposed as,
\bea
V_{n+1}^{S A_1\cdots A_{n} } = { V_{2}^{SA_{n}}} V_{n}^{S A_1\cdots A_{n-1}  }  \nonumber
\eea
from which a recurrence relation for a state $R^{(n)}$ is obtained,
\bea
R^{(n+1)} = \tr_{A_{n+1}} [ V_{2} (~ R^{(n)} \otimes  \rho^{A_{n+1}} ) ~ V_{2}^{\dagger} \widetilde{M}_{0}^{A_{n+1}}] \label{aeq:ares}
\eea
where $V_2$ acts on systems $S$ and  $A_{n+1}$. The elements also satisfy the following relations, 
\bea
R^{(n+1)}_{00} = \alpha R^{(n)}_{00} ~~\mathrm{and}~~ R^{(n+1)}_{11} = (1-\alpha)R^{(n)}_{11}~~\label{aeq:ar}
\eea
where $\alpha = (1-q)f + q(1-f)$. It is straightforward to see that $1-2\alpha<0$, from which $\alpha>1/2$. It follows that
\bea
\frac{R^{(n+1)}_{11}}{R^{(n+1)}_{00}} = \left( \frac{1 - \alpha}{ \alpha} \right) \frac{R^{(n )}_{11}}{R^{(n )}_{00}} =\cdots=\left( \frac{1 - \alpha}{ \alpha} \right)^{n+1} \frac{R^{(0 )}_{11}}{R^{(0 )}_{00}}. ~~~~~~\label{aeq:are}
\eea
Since $\alpha>1/2$, for all $\epsilon>0 $ one can find $n'\geq 1$ such that 
\bea
\frac{R^{(n')}_{11}}{R^{(n')}_{00}}  < \epsilon. \nonumber
%\xrightarrow{n\rightarrow \infty} 0\nonumber 
\eea
The fidelity in Eq. (\ref{aeq:afn}) converges to $1$ as $n$ increases. $\Box$

{\bf Remark 1.} {\it The protocol above extends to a non-uniform noise model in which each ancilla qubit $A_i$ has its own noisy state preparation $\rho^{A_i}$ and noisy measurement $\widetilde{M}^{A_i}_0$. } 
\begin{proof}    
Define
\begin{align}
f_i &:= \langle 0| \rho^{A_i} |0\rangle,
\qquad
q_i := \langle 1| \widetilde{M}^{A_i}_0 |1\rangle, \nonumber \\
\alpha_i &:= (1-q_i)f_i + q_i(1-f_i).
\label{aeq:alpha_i}    
\end{align}
Then the recurrence relation in Eq.~(\ref{aeq:ares}) becomes
\begin{align}
R^{(n+1)}
&=
\tr_{A_{n+1}}\!\left[
V_2
\left(
R^{(n)} \otimes \rho^{A_{n+1}}
\right)
V_2^\dagger\,
\widetilde{M}^{A_{n+1}}_0
\right],
\label{aeq:ares_nonuniform}
\end{align}
and the diagonal elements satisfy
\begin{align}
R_{00}^{(n+1)} &= \alpha_{n+1}\, R_{00}^{(n)},
&
R_{11}^{(n+1)} &= (1-\alpha_{n+1})\, R_{11}^{(n)}.
\label{aeq:ar_nonuniform}
\end{align}
Hence,
\begin{align}
\frac{R_{11}^{(n)}}{R_{00}^{(n)}}
&=
\left(
\prod_{i=1}^n \frac{1-\alpha_i}{\alpha_i}
\right)
\frac{R_{11}^{(0)}}{R_{00}^{(0)}}.
\label{aeq:ratio_nonuniform}
\end{align} 
Experimental parameters from calibrated devices satisfy that $f_i \ge 1/2 + a$ and $q_i \le 1/2 - b$ for all $i$ and for some $a,~b > 0$. For $\alpha_i = f_i(1-q_i) + q_i(1-f_i)$, we have $\alpha_i - \tfrac{1}{2} = 2(f_i-\tfrac{1}{2})(\tfrac{1}{2}-q_i) \ge 2ab >0 $. Therefore, we choose $\delta = 2ab$ so that $\alpha_i \ge \tfrac{1}{2} + \delta$ for all $i$. Then, it holds that, 
\begin{align}
\alpha_i &\ge \frac{1}{2}+\delta
\qquad
\text{for all}
\quad i\ge 1,
\label{aeq:uniform_margin}
\end{align}
then
\begin{align}
\frac{1-\alpha_i}{\alpha_i}
&\le
\frac{\tfrac{1}{2}-\delta}{\tfrac{1}{2}+\delta}
<1
\qquad
\text{for all}
\quad i\ge 1.
\end{align}
so that the product in Eq.~(\ref{aeq:ratio_nonuniform}) converges to $0$ exponentially fast. Consequently, the fidelity $f^{(n)}$ in Eq.~(\ref{aeq:afn}) still converges to $1$.
In particular, this holds whenever all ancilla preparations and measurements are better than random: $f_i>1/2$ and $q_i<1/2$ for all $i$.
\end{proof}

\section{Purification of Noisy Measurements }
\label{app:b}

{\it Summary of the protocol.} Outcomes of noiseless measurements can be distilled with noisy state preparations $\rho$ and noisy measurements, see also Fig. \ref{fig:ascenario}. To purify a noisy measurement on a system state $\sigma$, $m$ ancillary qubits are initialized in noisy preparations $\rho^{\otimes m}$. A collective CNOT $V_m$ is applied to the states $\sigma\otimes \rho^{\otimes m}$, and then noisy measurements are performed on the $(m+1)$ qubits individually. The outcomes are accepted only when they are identical, either $0^{m+1}$ or $1^{m+1}$. 

A noisy measurement can be generally described as follows, 
\bea
\widetilde{M}_k   = (1-q) M_k + q M_{k\oplus 1}~~\mathrm{for}~~ k=0,1, \label{aeq:nm}
\eea
where $q$ denotes a noise fraction. Otherwise, one can admix two instances, outcomes from a noisy measurement and the other obtained after applying a Pauli $Z$ so that a POVM element is given as above, i.e.,
\bea
\frac{1}{2} \left( \widetilde{M}_k + Z \widetilde{M}_k Z \right) = (1-q) M_k + q M_{ k\oplus 1}~~\mathrm{for}~~ k=0,1. \nonumber
\eea 
One can achieve the diagonalization by randomly admixing two probabilities, one from a noisy measurement and the other from a measurement after a Pauli gate $Z$.

Once outcomes are identical and thus accepted. We write that $\widetilde{N}_{k}^{(m)}$ is a POVM element giving outcome $k$ in the first register, given that $m+1$ outcomes are identical. The noisy fraction for the POVM element is given by $q^{(m)} = \langle \bar{k}| \widetilde{N}_{k}^{(m)} | \bar{k}\rangle$, where $\bar{k} = k \oplus 1$.\\

{\it {\bf Proposition 2.} For all $\epsilon>0$, there exists $N\geq 1$ such that $q^{(m)}<\epsilon$ whenever $m>N$.}\\

{\it Proof.}  A noisy POVM element can be purified with noisy SPAM as follows. The probability of having identical outcomes  $k^{m+1}$ for $k\in\{0,1\}$ is given by, for a state $\sigma$ on a system $S$,
\bea
p(k^{m+1})= %\sum_{k=0,1} 
\tr [V_{m+1} ~\sigma^S \otimes \rho^{\otimes m} ~V_{m+1}^{\dagger} ~ \M_{k}^{\otimes m+1} ]. \nonumber
\eea
From above, we write by $p^{(m)}(k)$ a probability of having an outcome $k$ on a system once noisy measurements on a system and $m$ target qubits give identical outcomes, 
\bea
p^{(m)} (k) = \frac{1}{p_{s}^{(m+1)}} \tr [ \sigma \M_{k}^{(m)}] \label{aeq:pum} 
\eea
where
\bea
\M_{k}^{(m)} & = & \tr_{ A_1\cdots A_m} [    \rho^{\otimes m} V_{m}^{\dagger} ~ M_{k}^S   \bigotimes_{i=1}^m \M_{k}^{A_i} ~ V_m ]\nonumber \\ 
~\mathrm{and}~ &~& {p_{s}^{(m+1)}} = p(0^{m+1})+p(1^{m+1}). \label{aeq:sup}
\eea 
The decomposition of a collective CNOT gate applies to derive a recurrence relation for a POVM element $\M_{k}^{(m)}$
\bea
\M_{k}^{(m+1)} & = & \tr_{ A_{m+1} } [ \rho^{A_{m+1}}    V_{2}^{  \dagger}   ~    \M_{k}^{(m) } \otimes  \M_{k}^{ A_{m+1} } ~ V_{2}^{ } ] ~~~~\label{aeq:rem}
\eea
where $V_2$ acts on systems $S$ and $A_{m+1}$. We solve the recurrence relation with the initial condition in Eq. (\ref{aeq:nm}) and have 
\bea
\M_{k}^{(m)} = \alpha^m(1-q) |k\rangle\langle k| +  (1-\alpha)^m q |\bar{k}\rangle \langle \bar{k}|, \nonumber
\eea
where $\alpha = f(1-q) + (1-f)q >1/2$. The success probability is also computed, $p_{s}^{(m+1)} =  \alpha^{m} (1-q) + (1-\alpha)^{m} q$. 
A purified POVM element in Eq. (\ref{aeq:pum}) corresponds to, 
\bea
%p^{(m)} (k) = \tr[ \sigma N_{k}^{(m)} ] ~\mathrm{where}~ 
\N_{k}^{(m)} =\frac{1}{p_{s}^{(m+1)} } \M_{k}^{(m)}. \label{aeq:pm}
\eea
The noise fraction is obtained as,
\bea
q^{(m)} = \left[1 + \left( \frac{\alpha}{1-\alpha} \right)^{m} \left(\frac{1-q}{q}\right)  \right]^{-1}. \nonumber
\eea
Since $\alpha>1/2$, for all $\epsilon>0$ there exists $m\geq 1$ such that $q^{(m)}<\epsilon$. Hence, it is shown that a noise fraction can be arbitrarily suppressed. $\Box$

{\bf Remark 2.} {\it  The proof above also extends to a non-uniform noise model in which each ancilla qubit $A_i$ has its own noisy state preparation $\rho^{A_i}$ and noisy measurement $\widetilde{M}^{A_i}_k$.}
\begin{proof}
Define
\begin{align}
f_i &:= \langle 0| \rho^{A_i} |0\rangle,
\qquad
q_i := \langle \bar{k}| \widetilde{M}^{A_i}_k |\bar{k}\rangle, \nonumber \\
\alpha_i &:= (1-q_i)f_i + q_i(1-f_i).
\label{aeq:alpha_i_meas}    
\end{align}
Then the recurrence relation in Eq.~(\ref{aeq:rem}) becomes
\begin{align}
\M_{k}^{(m+1)}
&=
\tr_{A_{m+1}}
\left[
\rho^{A_{m+1}}
V_{2}^{\dagger}
\left(
\M_{k}^{(m)} \otimes \M_{k}^{A_{m+1}}
\right)
V_{2}
\right],
\label{aeq:rem_nonuniform}
\end{align}
where $V_2$ acts on systems $S$ and $A_{m+1}$.
Writing
\begin{align}
r_0^{(m)} &:= \langle k| \M_k^{(m)} |k\rangle,
&
r_1^{(m)} &:= \langle \bar{k}| \M_k^{(m)} |\bar{k}\rangle,
\end{align}
one finds
\begin{align}
r_0^{(m+1)} &= \alpha_{m+1}\, r_0^{(m)},
&
r_1^{(m+1)} &= (1-\alpha_{m+1})\, r_1^{(m)}.
\label{aeq:r_nonuniform}
\end{align}
Hence,
\begin{align}
\frac{r_1^{(m)}}{r_0^{(m)}}
&=
\left(
\prod_{i=1}^m \frac{1-\alpha_i}{\alpha_i}
\right)
\frac{r_1^{(0)}}{r_0^{(0)}}.
\label{aeq:ratio_meas_nonuniform}
\end{align}
Therefore, if there exists $\delta>0$ such that
\begin{align}
\alpha_i &\ge \frac{1}{2}+\delta
\qquad
\text{for all}
\quad i\ge 1,
\label{aeq:uniform_margin_meas}
\end{align}
then the ratio in Eq.~(\ref{aeq:ratio_meas_nonuniform}) converges to $0$ exponentially fast. Since the noise fraction of the purified POVM element $\N_k^{(m)}$ is determined by the ratio $r_1^{(m)}/r_0^{(m)}$, it follows that $q^{(m)} \to 0$ as $m$ increases.
\end{proof}

\section{ Resources for Purifying Noisy SPAM }
\label{app:c}

In Table \ref{tab:1}, the purification with $n$ additional qubits and the success probability are computed in realistic scenarios, $q\in[0.05, 0.25]$, assuming errors are balanced, $1-f=q$. For instance, for $n=0$, without the purification, no post-selection is performed, $p_{s}^{(0)}=1$. 
% \begin{itemize}
% \item 
For SPAM errors less than $10\%$, in which $1-f=q=0.05$, it holds $T^{(1)}<10^{-2} $ with one additional qubit. The success probability is given as $p_{s}^{(1)} =0.86$. 
% \item 
For a higher error rate $1-f=q=0.1$, two additional qubits can be used to suppress an error up to $5\times 10^{-3}$ with the success probability $p_{s}^{(2)}=0.6$. 
% \end{itemize}
The purification with one additional qubit shows a success probability larger than $1/2$ in a realistic scenario.

\begin{table*}[t]
\caption{ The purification of noisy measurements with noisy SPAM is demonstrated.} 
\label{tab:1}
\centering
\begin{tabular}{|c|c|c||c|c||c|c|}
\hline
error rate($q$)/ purification(n) & $~~T^{(0)}~~$ & $~~p_{s}^{(0)}~~$ & $~~T^{(1)}~~$ & $~~p_{s}^{(1)}~~$ & $~~ T^{(2)}~~$ & $~~p_{s}^{(2)}~~$ \\
\hline \hline
$1-f=q=0.25$ & \makecell{$0.25$}& \makecell{$1$} 
& \makecell{$0.167$} &  \makecell{$0.562$} 
& \makecell{$0.107$} & \makecell{$0.328$} \\
\hline
$1-f=q=0.2 $ &  \makecell{$0.2$} & \makecell{$1$} 
& \makecell{$0.105$ } & \makecell{$0.608$} 
& \makecell{$0.052$ }  & \makecell{$0.390$} \\
\hline
$1-f=q=0.15 $ & \makecell{$0.15$} & \makecell{$1$} 
& \makecell{$0.057$ } & \makecell{$0.672$} 
& \makecell{$0.020$}  & \makecell{$0.482$} \\
\hline
$1-f=q=0.1 $ & \makecell{$0.1$} & \makecell{$1$} 
& \makecell{$0.024$ } & \makecell{$0.756$} 
& \makecell{$0.005$}  & \makecell{$0.608$} \\
\hline
$1-f=q=0.05 $ & \makecell{$0.05$ } & \makecell{$1$} 
& \makecell{$0.005$} & \makecell{$0.864$} 
& \makecell{$0.001$}  & \makecell{$0.779$} \\
\hline
\end{tabular}
\end{table*}

\section{Purification of SPAM with Non-uniform Noise}
\label{app:d}

Tables~\ref{tab:nonuniform_moderate} and~\ref{tab:nonuniform_strong} demonstrate the purification of noisy SPAM when noise parameters vary across qubits, as established in Remarks~1 and~2. The convergence requires only that each ancilla satisfies $\alpha_i = (1-q_i)f_i + q_i(1-f_i) > 1/2$. Table~\ref{tab:nonuniform_moderate} considers moderate noise; the results are comparable to the uniform case in Table~\ref{tab:1}. Table~\ref{tab:nonuniform_strong} considers a more demanding scenario in which the first ancilla has a preparation fidelity of $f_1 = 0.85$ and a measurement error of $q_1 = 0.12$. The protocol still converges, with lower success probabilities reflecting the reduced filtering power of the weaker ancilla.
\begin{table}[t]
\caption{ 
Purification of noisy SPAM with non-uniform moderate noise.
The system and ancilla qubits have parameters
$(f_S, q_S) = (0.97, 0.05)$,
$(f_1, q_1) = (0.94, 0.02)$,
$(f_2, q_2) = (0.98, 0.08)$,
$(f_3, q_3) = (0.96, 0.07)$.
}
\label{tab:nonuniform_moderate}
\centering
\begin{tabular}{c|cc|cc}
\hline\hline
 & \multicolumn{2}{c|}{State preparation}
 & \multicolumn{2}{c}{Measurement}\\
$n$ $(m)$
 & $f^{(n)}$ & $p_{\mathrm{succ}}^{(n)}$
 & $q^{(m)}$ & $p_{s}^{(m+1)}$\\
\hline
0 & 0.9700  & 1     & 0.0500   & 1     \\
1 & 0.9974  & 0.897 & $4.408\times 10^{-3}$ & 0.880 \\
2 & 0.9997 & 0.808 & $4.743\times 10^{-4}$ & 0.792 \\
3 & 0.99997 & 0.724 & $5.532\times 10^{-5}$ & 0.709 \\
\hline\hline
\end{tabular}
\end{table}

\begin{table}[t]
\caption{  
Purification of noisy SPAM with non-uniform strong noise.
The system and ancilla qubits have parameters
$(f_S, q_S) = (0.92, 0.07)$,
$(f_1, q_1) = (0.85, 0.12)$,
$(f_2, q_2) = (0.93, 0.11)$,
$(f_3, q_3) = (0.94, 0.09)$.
}
\label{tab:nonuniform_strong}
\centering
\begin{tabular}{c|cc|cc}
\hline\hline
 & \multicolumn{2}{c|}{State preparation}
 & \multicolumn{2}{c}{Measurement}\\
$n$ $(m)$
 & $f^{(n)}$ & $p_{\mathrm{succ}}^{(n)}$
 & $q^{(m)}$ & $p_{s}^{(m+1)}$\\
\hline
0 & 0.9200  & 1     & 0.0700   & 1     \\
1 & 0.9741  & 0.723 & $2.248 \times 10^{-2}$  & 0.7288 \\
2 & 0.9948 & 0.592& $4.510\times 10^{-3}$ & 0.598 \\
3 & 0.9998 & 0.579 & $7.321\times 10^{-4}$ & 0.513 \\
\hline\hline
\end{tabular}
\end{table}

\section{Purifying SPAM errors with noisy gates}
\label{app:e}

%So far, we have shown that purification works with the given resources of noisy SPAM for both cases of preparation and measurements. Namely, it suffices to work with noisy resources of SPAM {\it per se} to cleanse the errors from noisy SPAM. One can compare this to error correction, where noisy gates do not work to correct gate errors. Furthermore, quantum gates and SPAM working for diagnosing gate errors should contain sufficiently small error rates toward fault-tolerant computation \cite{doi:10.1137/S0097539799359385, 548464, doi:10.1126/science.279.5349.342}. 

%We observe the fact that CNOT gates in the purification protocol are noiseless. 

For the full generality, we consider the purification protocol when CNOT gates are noisy \cite{PhysRevLett.83.4200}, for which a noise model is,
\bea
V_2 (~\cdot~) V_{2}^{\dagger}\mapsto (1- \epsilon) V_2 (~\cdot~) V_{2}^{\dagger} +   \epsilon~\tr[~\cdot~] \frac{\I}{2} \otimes \frac{\I}{2},~ \label{aeq:noisecnot}
\eea
where $\epsilon$ denotes a noise fraction.

\subsection{Purifying noisy SPAM }
Applying the purification protocol when CNOT gates are noisy, one can find a recurrence relation for a state in Eq. (\ref{aeq:auns}),
\begin{align}
R_{00}^{(n+1)} &= (1-\epsilon) \alpha R_{00}^{(n)}  + \frac{\epsilon}{4} (R_{00}^{(n)}+R_{11}^{(n)}),~\mathrm{and} \nonumber\\
R_{11}^{(n+1)} &= (1-\epsilon) (1-\alpha) R_{11}^{(n)}  + \frac{\epsilon}{4} (R_{00}^{(n)}+R_{11}^{(n)}). \label{aeq:ara}
\end{align}
Note also that noisy SPAM are applied in the protocol. To see the convergence of a resulting state, let us compute the ratio from Eq. (\ref{aeq:ara}),
\bea
g &=& \lim_{n\rightarrow \infty} \frac{R_{11}^{(n)}}{R_{00}^{(n)}} =  \sqrt{D^2 +1} -D, \nonumber \\
&& \mathrm{where} ~~ D = 2(2f -1)(1-2q)\frac{ (1-\epsilon)}{\epsilon}. \label{aeq:ad} 
\eea
Then, the purification with $n$ target qubits leads to a convergent state up to a fidelity in the following, 
\bea
%\lim_{n\rightarrow \infty} 
f_{\epsilon}^{(n)} = \frac{R_{00}^{(n)} }{\tr R^{(n)}  } ~~  \xrightarrow{ n \rightarrow \infty }~~ \frac{1}{1-D+\sqrt{D^2 +1}}<1.  ~~\label{aeq:fep}
\eea
Thus, noiseless state preparation can be approximated with an error $1-f_{\epsilon}^{(n)}$. 

Similarly, for the purification of noisy measurements, the recurrence relation for a POVM element in Eq. (\ref{aeq:rem}) is obtained,
\bea
\langle k| {\M_k^{(m+1)}} |k\rangle & = & (1-\epsilon) \alpha   \langle k| {\M_k^{(m  )}} | k\rangle  + \frac{\epsilon}{4} \tr[\M_k^{(m )}], ~\mathrm{and}\nonumber \\
\langle \Bar{k} | {\M_k^{(m+1)}} | \Bar{k}\rangle &=& (1-\epsilon)(1-\alpha) \langle \Bar{k}| {\M_k^{(m )}} |\Bar{k}\rangle   + \frac{\epsilon}{4} \tr[\M_k^{(m )}]. \nonumber
\eea
Then, a noise fraction of a purified POVM element depends on both the number of target qubits $m$ and a noise fraction $\epsilon$; the noise fraction is also convergent,
\bea
q^{(m)}_{\epsilon} = \langle \bar{k} | \N_{k}^{(m,\epsilon)} | \bar{k} \rangle  ~~  \xrightarrow{ m \rightarrow \infty }~~ \frac{1}{1+D+\sqrt{D^2 +1}},   ~ \label{aeq:mep}
\eea
which is strictly positive, thus achieving a noiseless POVM element up to an error $q^{(m)}_{\epsilon}$. One can notice that a noise fraction above is equal to $1-f_{\epsilon}^{(m)}$ in Eq. (\ref{aeq:fep}).

\begin{figure*}[t]
    \centering
    \includegraphics[width=1\textwidth]{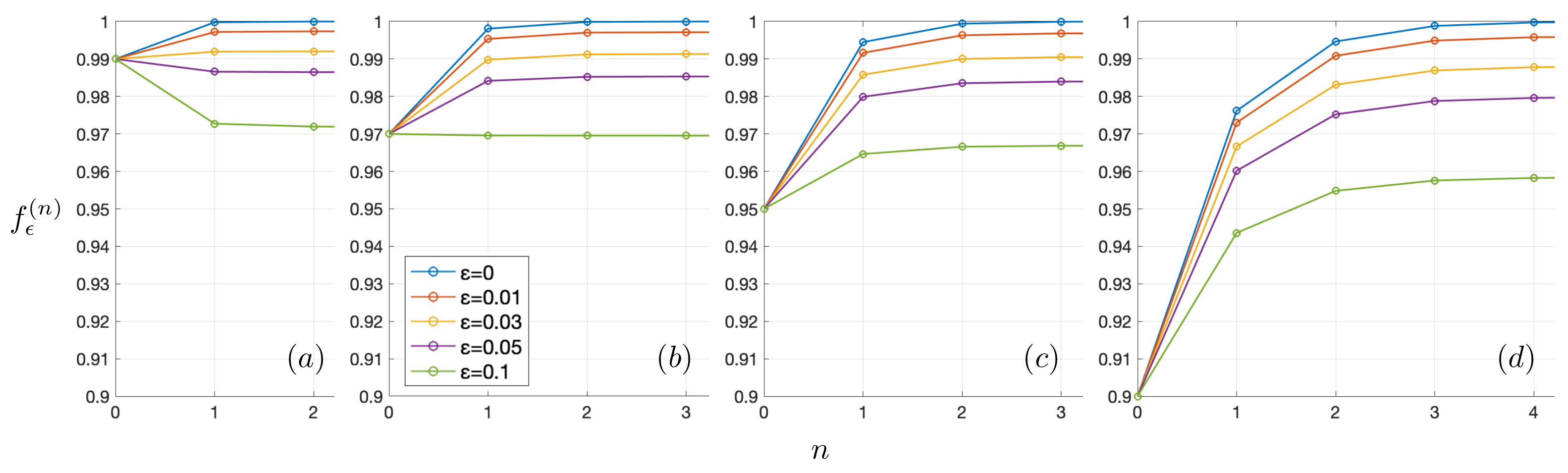}
    \caption{The purification of noisy initialization is demonstrated when a CNOT gate contains an error rate $\epsilon$ in Eq. (\ref{aeq:noisecnot}) and errors are balanced, $1-f= q$. An initial fidelity $f_{\epsilon}^{{(0)}}$ of state preparation may increase according to $n$ noisy target qubits. The convergence is $f_{\epsilon}^{(\infty)}$ in Eq. (\ref{aeq:fep}). For an initial fidelity $f=0.99$ in (a),  a single target qubit suffices to achieve noiseless preparation. A noisy CNOT gate with an error rate $\epsilon > 0.0374$ cannot be used to increase the fidelity; see Table \ref{e}. For $f=0.97$ and $f=0.95$ in (b) and (c), respectively, two or three noisy target qubits suffice to reach noiseless state preparation. Four target qubits are needed for $f=0.9$ in (d). For all cases, the purification protocol is efficient, and a few noisy target qubits are cost-effective.}
    \label{fig:agraph}
\end{figure*}

\subsection{Feasibility: Error Suppression in Realistic Scenarios}

From realistic constraints, let us consider balanced error rates $\epsilon=1-f=q=0.05$. The purification achieves a fidelity up to $0.984$ with two target qubits, i.e., the protocol can purify noisy state preparation from $0.95$ up to $0.984$. The success probability with two target qubits is given by {$p_{\rm succ}^{(2)}=0.7357$}. For a noisy measurement, the protocol suppresses an error rate from $0.05$ up to $0.016$, see Fig \ref{fig:agraph} (c).

\section{Verification of SPAM errors}
\label{app:g}

\begin{figure}[h]
    \centering
    \includegraphics[width=0.4\textwidth]{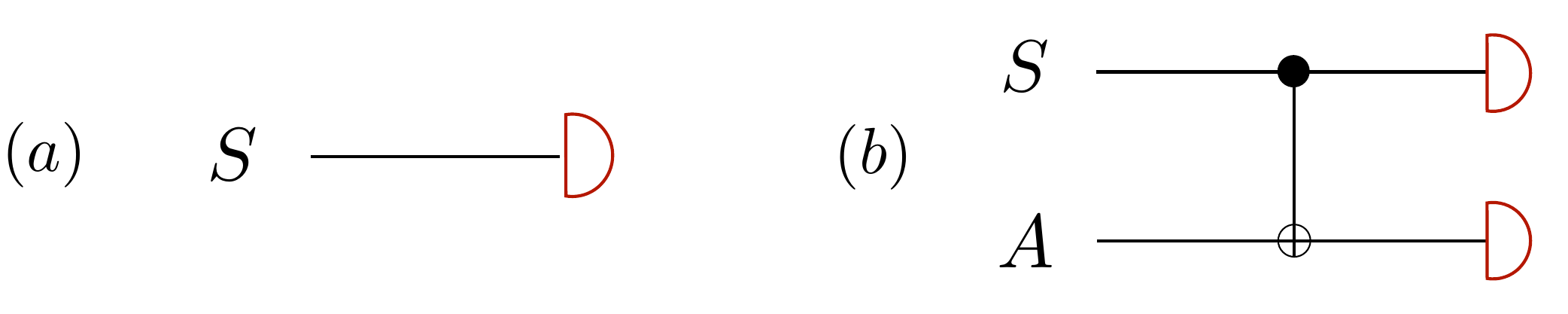}
    \caption{ (a) Single-copy level SPAM cannot verify errors of preparation and measurements. (b) Measurement outcomes can verify error rates in preparation, measurement, and a CNOT gate.}
    \label{fig:spam}
\end{figure}

The purification condition in Eq. (\ref{aeq:pc}) tells that it is necessary to learn error rates in SPAM, $1-f$ and $q$, to find if a purification works when a CNOT gate is noisy $\epsilon>0$. These error rates can be verified from measurements on two qubits, see Fig. \ref{fig:spam}. For noisy preparations with a fidelity $f$, a noisy CNOT gate in Eq. (\ref{aeq:noisecnot}) with an error rate $\epsilon$ is applied and noisy measurements with a noise fraction $q$ are performed. Probabilities $p(ij)$ for outcomes $i,j\in \{0,1 \}$ are given according to the error rates, 
\begin{align*}
p(00)\! & =\! (1-\epsilon)\big[f^2 (1-q)^2 \!+\! (1-f)^2 q^2 \!+\! f(1-f)(1-q)q 
\\ &\phantom{=} \!+\! (1-f)^2 q(1-q) \!+\! (1-f)fq^2\big] + \epsilon/4, \\
p(01)\! &=\! (1-\epsilon)\big[f^2 (1-q)q \!+\! (1-f)^2 q^2 \!+\! f(1-f)(1-q)^2 
\\ &\phantom{=} \!+\! (1-f)^2 q^2 \!+\! (1-f)fq(1-q)\big] + \epsilon/4, \\
p(10)\! &=\! (1-\epsilon)\big[f^2 q(1-q) \!+\! (1-f)^2 (1-q)q \!+\! f(1-f)q^2 
\\ &\phantom{=} \!+\! (1-f)^2 (1-q)^2 \!+\! (1-f)f(1-q)q\big] + \epsilon/4, ~ \mathrm{and} \\
p(11)\! &=\! (1-\epsilon)\big[f^2 q^2 \!+\! (1-f)^2 (1-q)^2 \!+\! f(1-f) q(1-q) 
\\ &\phantom{=} \!+\! (1-f)^2 (1-q)q \!+\! (1-f)f(1-q)^2\big] + \epsilon/4.
\end{align*}  
Hence, all noise parameters of SPAM and a noisy CNOT gate can be verified by solving equations above. 

In Table \ref{table:Sim}, the verification of parameters $f^{(0)}$, $q$, and $\epsilon$ from probabilities $p(ij)$ for $i,j=0,1$ is demonstrated. Once error rates are verified from probabilities of measurement outcomes, one can find if the purification condition, $f < f_{\epsilon}^{(1)}$, is fulfilled and may estimate the number of target qubits required for the purification. 

In Case 1, a CNOT gate is noiseless; an error rate of $10^{-3}$ is achieved with three additional target qubits. In Cases $2$, $3$, and $4$, a CNOT gate is noisy and the state preparation is purified up to $f_{\epsilon}^{(\infty)}$; a few qubits are cost-effective. In Case 5, a CNOT gate is too noisy to purify noisy state preparation: the condition $f < f_{\epsilon}^{(1)}$ is not fulfilled.

\begin{table}[h]
%\centering
\caption{ Critical error rates $\epsilon_c$ of a CNOT gate for purifying noisy SPAM are shown. }
\label{e}
\begin{tabular}{c||cccccccc}
\hline
\hline
~$1-f$ ~&& 0~& 0.01 ~&0.03~ & 0.05~ & 0.07~ & 0.1  \\  \hline
~ $\epsilon_c$ ~&& 0~& 0.0374 ~ & 0.0986~ & 0.1460~ & 0.1830~ & 0.2236 \\\hline
 \hline
\end{tabular}
\end{table}

\setlength{\tabcolsep}{12pt}
\renewcommand{\arraystretch}{1.2}
\begin{table*}[t]
\caption{ The purification of state preparation is demonstrated with noisy SPAM verified from measurement outcomes. }
\label{table:Sim}
\begin{center}
\begin{tabular}{|c||c|c|c||c|c|c| |c|c|c|c|c|c|c|c|}
\hline
\textbf{} & \multicolumn{3}{|c||}{\textbf{Measurement outcomes}}& \multicolumn{3}{|c||}{\textbf{Verification}} & \multicolumn{4}{|c|}{\textbf{Purification with $n$ qubits}} \\
\cline{2-11} 
\textbf{Cases} & $p(01)$ & $p(10)$ & $p(11)$ & $f^{(0)}$ &  $q$ & $\epsilon$ & $f^{(1)}$ & $f^{(2)}$ & $f^{(3)}$ & $f^{(\infty)}$ \\
\hline
1 & $ 0.154 $ & $0.09$ & $0.09$ & $ 0.9$ & $0.1$ & $0$ & $0.976$ & $0.995$& $0.999$ & $1$ \\
\hline
2 & $ 0.1252 $ & $0.0540$ & $0.0896$ & $ 0.9$ & $0.05$ & $0.01$ & $0.979$ & $0.994$& $0.996$ & $0.996$  \\
\hline
3 & $ 0.0777 $ & $0.0530$ & $0.0366$ & $ 0.97$ & $0.05$ & $0.03$ & $0.989$ & $0.991$& $0.991$ & $0.991$  \\
\hline
4 & $ 0.0961 $ & $0.0576$ & $0.0576$ & $ 0.95$ & $0.05$ & $0.05$ & $0.980$ & $0.983$& $0.984$ & $0.984$  \\
\hline
5 & $ 0.0754 $ & $0.0674$ & $0.0357$ & $ 0.99$ & $0.05$ & $0.1$ & $0.99$ & $0.971$& $0.970$ & $0.970$  \\
\hline
\end{tabular}
\end{center}
\end{table*}

\section{ Applications to quantum networks }  
\label{app:h}

Distilling and swapping entanglement are essential to realize a quantum network. Entanglement distillation enables two parties to share an entangled bit (ebit) over a distance $L$, $|\phi^+\rangle = \frac{1}{\sqrt{2}} (|00\rangle +|11\rangle)$ from noisy entangled states. Once two parties Alice and Bob share ebits with a repeater, $|\phi^+\rangle^{AR_1}$ and $|\phi^+\rangle^{ R_2 B}$, a joint measurement on $R_1R_2$ in a repeater creates an ebit $|\phi^+\rangle^{AB}$ between Alice and Bob. Then, entanglement swapping allows two parties to share an ebit over a distance $2L$.

To see how measurements in entanglement distillation can be noisy, we recall that two parties firstly apply twirling to share copies of one-parameter states, called Werner states, 
\bea
\rho_F = F|\phi^+\rangle \langle \phi^+| + \frac{1-F}{3}\left( \I \otimes \I- |\phi^+\rangle \langle \phi^+|\right), \nonumber
\eea
which is entangled if and only if the fidelity satisfies $F>1/2$. After applying bilateral CNOT gates on two copies of a Werner state, two parties perform measurements in the second register; when measurement outcomes are identical, a resulting state in the first register is accepted. Let $F^{'}$ denote a resulting fidelity and we have $F^{'}>F$ whenever $F>L$ where $L=1/2$, i.e., an entangled Werner state can be distilled. If measurements in the second register are noisy, we have some threshold $L > 1/2$ \cite{PhysRevA.59.169}. 

\begin{figure*}[t]
    \centering
    \includegraphics[width=0.95\textwidth]{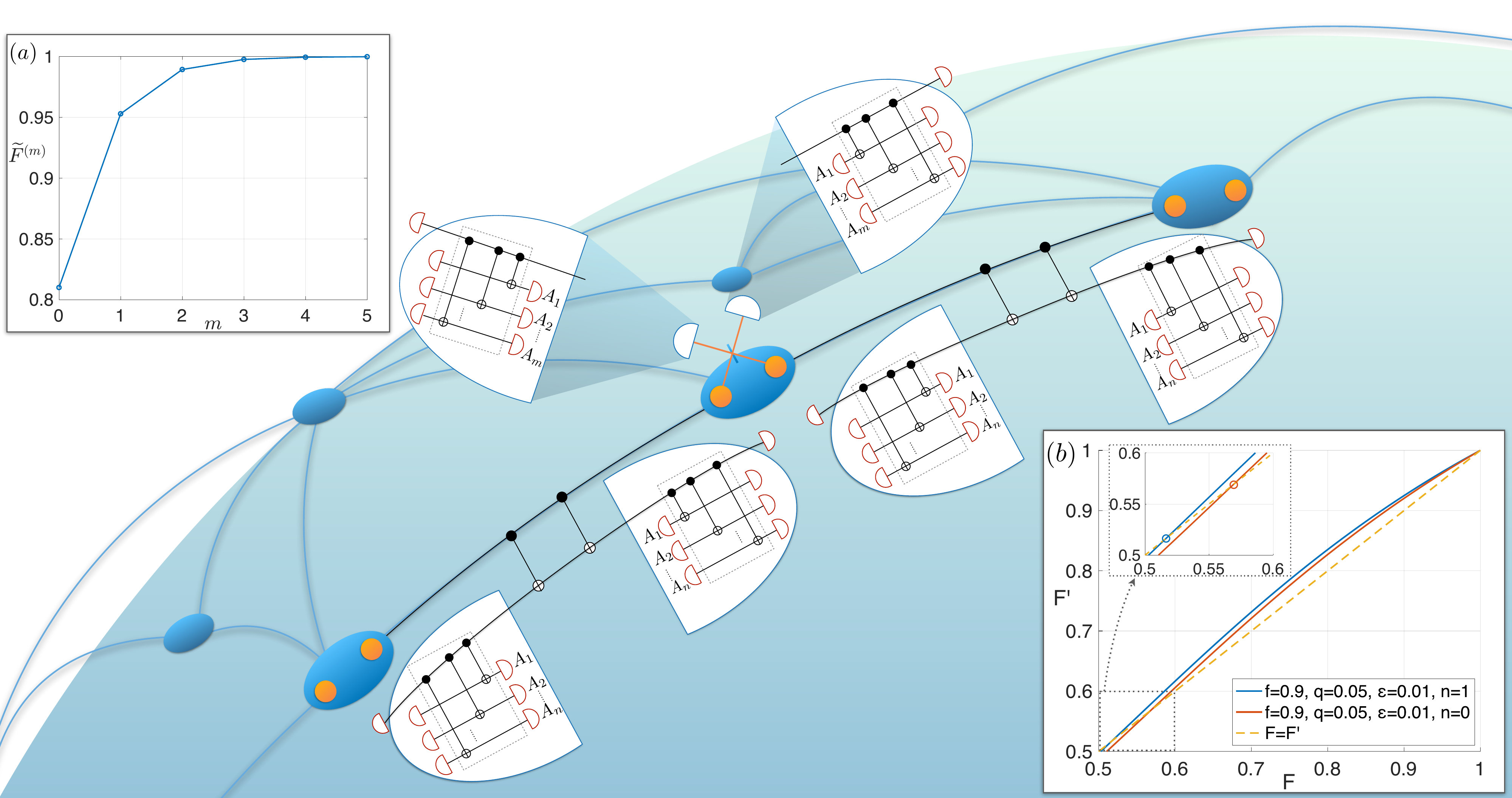}
    \caption{ Noisy measurements in a quantum network may appear in entanglement distillation and entanglement swapping in repeaters. In entanglement swapping in (a), a network results in a pure state with a sufficiently a high fidelity by purifying noisy measurements, for which three noisy qubits are cost-effective. In (b), noisy measurements in entanglement distillation can be purified by noisy SPAM; a lower bound for distillation improves from $0.569$ (red) to $0.517$ (blue) where the purification works with one noisy target qubit having a fidelity $0.9$, a measurement with an error rate $0.05$ and a noisy CNOT gate with noise fraction $0.05$. }
  \label{fig:network}
\end{figure*}

With the purification protocol for noisy SPAM, we now show that an entangled network can be achieved through entanglement distillation and entanglement swapping by purifying noisy resources for SPAM, see Fig. \ref{fig:network}. 

\subsection{ Distilling Entanglement with Noisy SPAM }
\label{subsec:ed}
 
In this section, we purify noisy measurements and show that entanglement can be distilled with noisy SPAM. The result resolves the problem pointed out in Ref. \cite{PhysRevA.59.169} that noisy measurements in entanglement distillation \cite{PhysRevLett.76.722} prevent parties from distilling entanglement from weakly entangled states. 

\bigskip
{\it  {\bf Proposition 3.}  A larger set of entangled states can be distilled by purifying noisy SPAM.}
\bigskip
 
{\it Proof.} Consider two nodes in a network with noisy resources for SPAM. Applying the protocol for purifying noisy measurements, we have a recurrence relation,
\bea
\M_{k}^{(n )} & = & \tr_{ A_{n} } [ \rho^{A_{n}}    V_{2}^{  \dagger}   ~    \M_{k}^{(n-1) } \otimes  \M_{k}^{ A_{n} } ~ V_{2}^{ } ] ~~\label{aeq:arem}
\eea
where noisy SPAM are only exploited. We also write by $r_0^{(n)} = \langle k| {\M_k^{(n)}} |k\rangle$ and $ r_1^{(n)} = \langle {\bar k}| {\M_k^{(n)}} | {\bar k}\rangle $. From the recurrence relation, we have
\bea
r_{0}^{(n+1)} & = &  (1-\epsilon) \alpha  r_{0}^{ (n) }  + \frac{\epsilon}{4}  ( r_{0}^{(n)} + r_{1}^{(n)} ) \nonumber \\
r_{1}^{(n+1)} & = & (1-\epsilon) (1-\alpha) r_{1}^{(n)}   + \frac{\epsilon}{4} (r_{0}^{(n)} + r_{1}^{(n)} ). \label{aeq:rr}
\eea
With the purification on noisy measurements, Alice and Bob apply the entanglement distillation protocol in Ref. \cite{PhysRevLett.76.722}. We have a resulting fidelity $F^{'}$, %see also Eq. (\ref{aeq:fidel})
\bea
F^{'} = \frac{F^2 +  \left( \frac{1-F}{3} \right)^2  + g^{{(n)}}(F) }{ F^2 + 2F \left( \frac{1-F}{3} \right) +  5 \left( \frac{1-F}{3} \right)^2+  4g^{{(n)}}(F)  } \label{aeq:FF}
\eea
where
\bea
g^{{(n)}}(F)  =  \left(   \frac{F(1-F)}{3} + \left( \frac{1-F}{3} \right)^2  \right) \frac{r_{odd}^{(n)}}{r_{even}^{(n)}} \nonumber
\eea
with $r_{odd}^{(n)} = 2 r_{0}^{(n)} r_{1}^{(n)}$ and $r_{even}^{(n)} = r_{0}^{(n)}r_{0}^{(n)} + r_{1}^{(n)}r_{1}^{(n)}$. 
From the relations, we compute
\bea
F^{'} - F = 8(F-\frac{1}{4}) (F- L^{(n)} ) (1-F ) \nonumber
\eea
and derive the distillability condition: $F> L^{(n)}$, where
\bea
L^{(n)}  = \frac{1}{2} \left(\frac{r_{0}^{(n)} + r_{1}^{(n)}  }{r_{0}^{(n)} - r_{1}^{(n)} } \right)^2. \nonumber%~~ \xrightarrow{n \rightarrow \infty} ~~ \frac{1}{2} \nonumber
\eea

On the one hand, suppose that CNOT gates in the purification are noiseless, $\epsilon=0$. We can solve recurrence relations in Eq. (\ref{aeq:rr}), $r_0^{(n)} = (1-q) \alpha^n$ and $ r_1^{(n)} = q (1-\alpha)^n$. It follows that 
\bea
L^{(n)} ~~ \xrightarrow{n \rightarrow \infty} ~~ \frac{1}{2} \label{aeq:conv}
\eea
meaning that an ebit can be distilled from all entangled two-qubit states. 

On the other hand, let us consider a case where a CNOT gate is noisy, $\epsilon>0$, for which we derive a relation 
\bea
\frac{r_{1}^{(n+1)}}{r_{0}^{(n+1)}} & = &  \frac{ (1-\epsilon) (1-\alpha) r_{1}^{(n)}   + \frac{\epsilon}{4} (r_{0}^{(n)} + r_{1}^{(n)} ) }{ (1-\epsilon) \alpha  r_{0}^{ (n) }  + \frac{\epsilon}{4}  ( r_{0}^{(n)} + r_{1}^{(n)} )} \nonumber \\
 & = &  \frac{ (1-\epsilon) (1-\alpha) \left( \frac{r_{1}^{(n)}}{r_{0}^{(n)}} \right)  + \frac{\epsilon}{4} \left( 1  + \left(\frac{r_{1}^{(n)}}{r_{0}^{(n)}} \right)\right)
  }{ (1-\epsilon) \alpha   + \frac{\epsilon}{4}  \left( 1  + \left( \frac{r_{1}^{(n)}}{r_{0}^{(n)}}\right) \right)}.~~~~~~  \label{aeq:ars} 
\eea
From the above, we compute the convergence,  
\bea
\frac{r_{1}^{(n)}}{r_{0}^{(n)}} \rightarrow d= \sqrt{D^2 +1} -D 
\eea
where 
\bea
d= \sqrt{D^2 +1} -D,~\mathrm{and}~ D = 2 (\frac{ 1-\epsilon}{\epsilon}) (2f-1) (1-2q).  \nonumber
\eea
Thus, the lower bound is convergent as follows, 
\bea
L^{(n)} ~~ \xrightarrow{n \rightarrow \infty} ~~   \frac{1}{2} \left( \frac{1 + d }{1 -d} \right)^2 \label{aeq:convv}
\eea
which is higher than the ultimate lower bound $1/2$. $\Box$

We note that the success probability in the distillation protocol, i.e., the probability that the copy in the first register is accepted, is given by
\begin{align}
p_{\rm succ}^{(n)}(F) &= \left( F^2 + 2F \left(\frac{1-F}{3}\right) + 5\left(\frac{1-F}{3}\right)^2 \right) r_{even}^{(n)} \nonumber
\\& \phantom{=} + \left(4F \left(\frac{1-F}{3}\right) + 4F \left(\frac{1-F}{3}\right)^2 \right) r_{odd}^{(n)}. \label{aeq:psucc}    
\end{align}
The success probability in Eq. (\ref{aeq:psucc}) takes two steps into account: one is the probability that the purification protocol with $n$ qubits is successful, and the other is the probability that the first copy is accepted in the distillation protocol.  

\begin{table*}[t]
\caption{  The number of copies of shared states $N_c$ for distilling an ebit is computed when a measurement error is $5\%$, see Eq. (\ref{aeq:ns}).  } 
\label{tab:fidelity}
\centering
\begin{tabular}{|c|c|c|c|c|c|}
\hline
$  $ & \makecell{ $N_c$  \\ $ p_{\rm succ}^{(0)} (F_0)$ } &  \makecell{ $N_c$  \\ $ p_{\rm succ}^{(1)}(F_0)$ } &  \makecell{ $N_c$  \\ $ p_{\rm succ}^{(2)} (F_0)$ } &  \makecell{ $N_c$  \\ $ p_{\rm succ}^{(3)} (F_0) $ } &   \makecell{ $N_c$  \\ $ p_{\rm succ}^{(4)} (F_0)$ } \\
\hline \hline
$F_0 = 0.6, 1-f=q=0.05$ %\vrule $~$ %\makecell{Required copies \\ \hline $p_s^{(1)}$ %} 
& undistillable & \makecell{$4.904 \times 10^{11}$ \\ $0.453$} & \makecell{$1.184\times 10^{13}$ \\ $0.369$} &  \makecell{$9.371\times 10^{14}$ \\ $0.302$} & \makecell{$7.557\times 10^{16}$ \\ $0.247$} \\
\hline
$F_0 = 0.7, 1-f=q=0.05$ & \makecell{$1.432\times 10^{12}$ \\ $0.646$} & \makecell{$1.323\times 10^{9}$ \\ $0.505$} & \makecell{$1.657\times 10^{10}$ \\ $0.412$} & \makecell{$5.985\times 10^{11}$ \\ $0.337$} & \makecell{$2.175\times 10^{13}$ \\ $0.276$}  \\
\hline
$F_0 = 0.8, 1-f=q=0.05$ & \makecell{$1.937\times 10^{9}$ \\ $0.718$} & \makecell{$2.494\times 10^{7}$ \\ $0.570$} & \makecell{$5.914\times 10^{8}$ \\ $0.466$} & \makecell{$1.438\times 10^{10}$ \\ $0.381$} & \makecell{$3.507\times 10^{11}$ \\ $0.312$} \\
\hline
$F_0 = 0.9, 1-f=q=0.05$ & \makecell{$1.429\times 10^{7}$ \\ $0.804 $} & \makecell{$6.627\times 10^5$ \\ $0.648$} & \makecell{$8.798\times 10^6$ \\ $0.530$} & \makecell{$1.178\times 10^{8}$ \\ $0.434$} & \makecell{$1.578\times 10^{9}$ \\ $0.356$} \\
\hline
\end{tabular}
\end{table*}

\subsection{ Optimizing Resources: Single Target Qubit is Optimal}

Although the convergence is shown in Eqs. (\ref{aeq:conv}) and (\ref{aeq:convv}), it is left to answer how large $n$ is needed for distilling ebits from a larger set of entangled states. With the constraint on a fidelity with an ebit higher than $0.999$, we estimate the number of additional qubits for distilling $1$ ebit in what follows. 

Note that each round of the distillation protocol makes a post-selection of $N p_{\rm succ}^{(n)}/2$ copies out of $N$ shared copies. Let $F_j$ denote a singlet fidelity after $j$ rounds: an initial fidelity is thus denoted by $F_0$ and a final one $F_J >0.999$.

Then, the critical number of copies $N_c$ having an initial fidelity $F_0$ for distilling $1$ ebit is given by 
\bea
N_c = \prod_{j=0}^{J-1}   \frac{2}{ p_{\rm succ}^{(n)}(F_j) },  \label{aeq:ns}
\eea
with the purification of noisy measurements with $n$ qubits. 

The results for realistic scenarios where preparation and measurement errors are $5\%$ are shown in Table \ref{tab:fidelity}. An initial fidelity $F_0$ is given for $[0.6,0.9]$. 
\begin{itemize}
\item $N_c$ is minimal when one additional qubits are exploited on both sides. Thus, one additional qubits on both sides are optimal for distilling entanglement by purifying noisy SPAM. 
\item The success probabilities when additional qubits are exploited for the purification are comparable, see e.g., $p_{\rm succ}^{(1)} (F_0)$ and $p_{\rm succ}^{(0)} (F_0)$. 
\end{itemize}

{ 
Notably, $N_c$ is not monotonic in $n$: the minimum occurs at $n = 1$ for all initial fidelities considered. This reflects a trade-off between measurement quality and success probability. Increasing $n$ further suppresses the noise fraction $q^{(n)}$, which lowers the distillation threshold $L^{(n)}$ and may reduce the number of distillation rounds $J$ required to reach the target fidelity. However, each additional ancilla also reduces the per-round success probability $p_{\rm succ}^{(n)}(F_j)$ by imposing an additional post-selection condition. At $n = 1$, the dominant measurement errors are already substantially suppressed (from $q = 0.05$ to $q^{(1)} \approx 0.005$), and the marginal improvement from further ancillae does not compensate for the reduced success probability. Thus, a single ancilla per party is optimal for distilling entanglement by purifying noisy SPAM under chosen parameters.
}

\subsection{Entanglement Swapping}
\label{subsec:es}

For ebits shared with a repeater $R=R_1R_2$, $|\phi^+\rangle^{AR_1}|\phi^+\rangle^{ R_2B}$, a noiseless Bell measurement is desired, where a projection onto an ebit $|\phi^+\rangle^{R_1R_2}$ enables two parties Alice and Bob to share an ebit. 
A successful projection for a state $\rho^{AR_1R_2B}$, denoted by a success probability $p_R$, can be realized in practice as follows,
\bea
p_{R}\rho^{AB} = \tr_{R_1R_2}[ \rho^{AR_1R_2B}  V_2 H^{R_1} |00\rangle \langle 00|^{R_1R_2 }   H^{R_1} V_{2}^{\dagger}] \label{aeq:sb}
\eea
since $|\phi^+\rangle = V_2 H^{R_1} |00\rangle$ where $H^{R_1}$ denotes a Hadamard transform on $R_1$ and $V_2$ acts on $R_1R_2$. Then, outcomes $00$ correspond to a successful projection in Eq. (\ref{aeq:sb}). 

Noisy measurements lead to  
\bea
\widetilde{p}_{R} \widetilde{\rho}^{AB}  =\tr_{R_1R_2}[ \rho^{AR_1R_2B}  V_2 H^{ R_1 }  \M_{0}\otimes \M_0 H^{ R_1 } V_{2}^{\dagger}],  \label{aeq:sig}
\eea
the result of which leads to a mixture of four Bell states between Alice and Bob.

\bigskip
{\it  {\bf Proposition 4.}  Entanglement swapping can be realized by noisy SPAM with the purification protocol.}
\bigskip

{\it\ Proof.} A purified measurement with $m$ noisy target qubits on $R_1$ and $R_2$, respectively, constructs POVM elements $\N_{0}^{(m)} \otimes \N_{0}^{(m)}$. The resulting state of Alice and Bob after applying the measurements on two ebits may be denoted by $\sigma^{(m)}$ and can be quantified by the fidelity, 
\bea
\widetilde{F}^{(m)} = \langle \phi^+| \sigma^{(m)} |\phi^+\rangle   = \left[ 1+ \left( \frac{1-\alpha}{\alpha} \right)^m \left(\frac{q}{1-q}\right)   \right]^{-2}\nonumber
\eea
Since $\alpha>1/2$, one can find $m\geq 1$ such that $\widetilde{F}^{(m)}$ is sufficiently close to $1$. $\Box$

Thus, noisy measurements distributing mixed states over a network are purified, which then enables a noiseless and large-size entangled network, see Fig. \ref{fig:network}.

\end{document}